\documentclass[apj]{emulateapj}
\usepackage{ulem}
\usepackage{amsmath}
\usepackage{graphicx}
\usepackage{subfigure}
\usepackage{multirow}
\usepackage{color}
\shorttitle{Interacting Cosmic Rays in M82}
\shortauthors{Yoast-Hull, Everett, Gallagher \& Zweibel}
\begin{document}
\title{Winds, Clumps, and Interacting Cosmic Rays In M82}

\author{Tova M. Yoast-Hull$^1$, John E. Everett$^{1,2,3,4,5}$, J. S. Gallagher III$^4$, and Ellen G. Zweibel$^{1,4,5}$}

\affil{$^1$Department of Physics, University of Wisconsin-Madison, WI, USA; email: {\tt yoasthull@wisc.edu}}
\affil{$^2$Center for Interdisciplinary Exploration and Research in Astrophysics, Northwestern University, IL, USA;}
\affil{$^3$Department of Physics \& Astronomy, Northwestern University, IL, USA;}
\affil{$^4$Department of Astronomy, University of Wisconsin-Madison, WI, USA; \\}
\affil{$^5$Center for Magnetic Self-Organization in Laboratory and Astrophysical Plasmas, University of Wisconsin-Madison, WI, USA \\}


\begin{abstract}
We construct a family of models for the evolution of energetic particles in the starburst galaxy M82 and compare them to observations to test the calorimeter assumption that all cosmic ray energy is radiated in the starburst region.  Assuming constant cosmic ray acceleration efficiency with Milky Way parameters, we calculate the cosmic-ray proton and primary and secondary electron/positron populations as a function of energy.  Cosmic rays are injected with Galactic energy distributions and electron-to-proton ratio via type II supernovae at the observed rate of  0.07~yr$^{-1}$.  From the cosmic ray spectra, we predict the radio synchrotron and $\gamma$-ray spectra.  To more accurately model the radio spectrum, we incorporate a multiphase interstellar medium in the starburst region of M82.  Our model interstellar medium is highly fragmented with compact dense molecular clouds and dense photoionized gas, both embedded in a hot, low density medium in overall pressure equilibrium.  The spectra predicted by this one-zone model are compared to the observed radio and $\gamma$-ray spectra of M82.  $\chi^{2}$ tests are used with radio and $\gamma$-ray observations and a range of model predictions to find the best-fit parameters.  The best-fit model yields constraints on key parameters in the starburst zone of M82, including a magnetic field strength of $\sim$250~$\mu$G and a wind advection speed in the range of 300-700~km~s$^{-1}$.  We find that M82 is a good electron calorimeter but not an ideal cosmic-ray proton calorimeter and discuss the implications of our results for the astrophysics of the far infrared-radio correlation in starburst galaxies.
\end{abstract}


\section{Introduction}

The coupling between star formation activity and interstellar gas dynamics is fundamental to galactic structure and evolution.  Magnetic fields and cosmic rays are a key part of that feedback.  In this paper, we fit radio and $\gamma$-ray spectra to study cosmic rays and magnetic fields in the archetypal starburst galaxy M82.

With the discovery of the far infrared-radio luminosity (FIR-radio) correlation it became clear that the energetic cosmic ray particle population in galaxies is closely related to the star formation process \citep{Helou85}.  Since far infrared emission from galaxies is primarily powered by dust absorption of photon radiation from young, massive stars and the 20-cm radio continuum by relativistic cosmic-ray electrons, the FIR-radio correlation requires physical connections to exist between star formation processes, cosmic ray production, and the physical state of the ISM.  These connections have now been further extended by $\gamma$-ray observations of galaxies obtained at TeV energies from the ground with VERITAS and HESS and from space at lower energies with {\it Fermi}.

Starburst galaxies are particularly interesting for testing models of the FIR-radio correlation in galaxies.  While electron losses in normal galaxies occur via synchrotron radiation in a generally diffuse ISM filling large volumes, the situation in starbursts is much more extreme.  That this correlation applies to starbursts, with their intense star formation, high pressure ISM, and large-scale galactic winds, as well as normal galaxies, is puzzling as the radio synchrotron emission depends on the electron energy density and the magnetic field strength \citep{Volk89}, as well as losses via galactic winds.

The observed validity of the FIR-radio correlation has led to the concept of calorimeter models for cosmic ray interactions in galaxies. The underlying principle of the calorimeter model is that the energy radiated by relativistic particles scales with the relativistic particle energy input. As cosmic rays are produced by supernovae, the total energy in cosmic rays follows from the supernova rate and thus from the star-formation rate (SFR).  A galaxy is then a calorimeter if essentially all the power that supernovae transfer to relativistic particles can be balanced by observable radiative losses.  If the calorimeter model applies to cosmic-ray electrons, then the synchrotron luminosity is proportional to supernova rate and thus the SFR, and the FIR-radio correlation is obtained. If it applies to cosmic-ray protons, then the $\gamma$-ray luminosity also depends only on the supernova rate. The actual situation is complicated by additional factors, including the effect of the magnetic fields on synchrotron radiative efficiencies, non-radiative losses such as ionization, and advection of particles and magnetic fields if a galactic wind is present.  The existence of such secondary factors that especially influence synchrotron luminosities make the existence of the FIR-radio correlation all the more remarkable.

Observations with HESS and VERITAS on the ground and {\it Fermi} in space are filling in the gap in measurements of high energy $\gamma$-ray photons from galaxies, including nearby starburst systems \citep[e.g.,][]{Acero09,Abramowski12,Ackermann12}.  Analysis of these observations provide unique insights into interactions between cosmic rays and the interstellar medium (ISM), which in turn lead to a better understanding of the distribution of feedback energy supplied by supernovae \citep[e.g.,][]{Paglione96,Torres04,delPozo09a,Lacki10,Paglione12}.  Further progress in assessing the high energy particle content of galaxies comes from radio measurements of the synchrotron emission produced by relativistic electrons. Radio observations also are advancing in terms of dynamic range, improved angular resolution, and sensitivity, especially at low frequencies.  Thus a combination of $\gamma$-ray and radio spectra can be analyzed to yield information on the high energy particle content and magnetic fields in galaxies \citep[e.g.,][]{Thompson06}.

We selected M82 for this initial study as this galaxy has the advantages of a well determined supernova rate, an extensively studied starburst region that includes estimates of the mass of interstellar gas, a carefully studied galactic wind, and measurements of the $\gamma$-ray and radio spectral energy distributions.  M82 therefore has been a target of choice in testing proton calorimeter models to determine the degree to which cosmic ray energy is deposited within the galaxy \citep[e.g.,][]{Thompson06,Persic08,delPozo09a,Lacki11,Paglione12}.

In this paper we present a one-zone model for calculating the electromagnetic power radiated by relativistic cosmic-ray protons and electrons in starburst galaxies.  Our approach is based on analytic calculations applied to a starburst zone, where, due to the compact nature of the region and high space density of supernovae, cosmic ray diffusion is not important.  We include physical processes associated with cosmic ray interactions with interstellar matter and synchrotron radiation produced by primary and secondary cosmic ray electrons.  This model also incorporates a multi-phase ISM in approximate pressure equilibrium and allows for the effects of galactic winds that advect particles out of the starburst zone.  Emergent fluxes are calculated for a model of M82 and fits to both the radio and $\gamma$-ray spectra are used to determine key parameters that are not otherwise directly constrained by observations of M82.  Our approach has the advantage of computational simplicity while retaining the key physics.  This allows us to efficiently search a wide range of parameters as discussed in \S 2 to find models that best fit the observed radio and $\gamma$-ray spectra.

Our approach to modeling the starburst zone builds on that in previous models \citep[e.g.,][]{delPozo09a, Paglione12} and explores the effects of a wind on the system in detail.  Unlike previous models, we treat the starburst core and the halo separately when dealing with radio emission from the galaxy as there is an expected turn-over of the radio flux at low frequencies in the starburst core \citep{Adebahr}.  Additionally, we use a fully consistent energy loss lifetime to test the calorimeter model while previous explorations of the calorimeter model are limited to semi-empirical formulae \citep{Ackermann12}.

The next section describes the properties of M82.  Section 3 shows how the population of energetic particles was computed.  Section 4 contains the details for how the model was applied and describes our results and the landscape of parameter space.  In Section 5, we discuss the calorimeter model and implications for the FIR-radio correlation, and in Section 6, we present concluding remarks.

\begin{center}
\begin{deluxetable}{llc}
%
\tablecaption{YEGZ Model Parameters}
\tablewidth{0pt}
\tablehead{
\colhead{Physical Parameters} & \colhead{Values Adopted} & \colhead{Reference}
}
\startdata
Radius SB & 200 pc & 1 \\
Scale Height SB & 100 pc & 1 \\
Distance & 3.9 Mpc & 2 \\
Molecular Gas Mass & 2-4$\times$10$^{8}$ M$_{\odot}$ & 3,4 \\
Ionized Gas Mass & 8$\times$10$^{6}$ M$_{\odot}$ & 5 \\
Ionized Gas Temperature & 8000 K & 5 \\
Hot Gas Temperature & 6$\times$10$^{6}$ K & 6 \\
Hot Gas Density\tablenotemark{a} & 0.33 cm$^{-3}$ & 6 \\
Average ISM Density\tablenotemark{a} & 280-550 cm$^{-3}$ & \\
IR Luminosity & 4$\times$10$^{10}$ L$_{\odot}$ & 7 \\
Radiation Field Energy Density\tablenotemark{a} & 1000 eV cm$^{-3}$ & \\
SN Explosion Rate & 0.07 yr$^{-1}$ & 8 \\
SN Explosion Energy\tablenotemark{b} & 10$^{51}$ ergs & \\
SN Energy Transferred to CR\tablenotemark{b} & 10\% & \\
Ratio of Primary Protons & 50 & \\
~~~to Electrons (N$_{p}$/N$_{e}$) & & \\
Slope of Primary CR & 2.1/2.2 & \\
~~~Source Function & & \\
\enddata
%
%
\tablenotetext{a}{Derived from above parameters}
\tablenotetext{b}{Excludes neutrino energy}
\tablerefs{
[1] \cite{Forster03a}; [2] \cite{Sakai99}; [3] \cite{Naylor10}; [4] \cite{Wild92}; [5] \cite{Forster01}; [6] \cite{Stevens03}; [7] \cite{Rice88}; [8] \cite{Fenech08};
}
\end{deluxetable}
\end{center}
%


\section{Properties of M82}


\subsection{Structure}

M82 is a nearby \citep[D$=$3.9~Mpc;][]{Sakai99} and well-studied example of a central starburst occurring in a moderate mass disk galaxy \citep[e.g.,][]{O'Connell78}.  The galaxy is nearly edge on, so that much of the starburst zone is optically obscured by the dusty ISM within the disk.  This orientation, however, allows the galactic wind to be clearly seen in wavelengths extending from the x-rays to the radio. M82 is a member of the M81 galaxy group and is connected to M81 and its surroundings by tidal debris that is readily visible in H~\textsc{i} 21-cm line maps \citep{Yun94,Chynoweth08}. This structure is a result of a close passage about 0.2~Gyr ago between M82 and M81 that is the probable trigger of the M82 starburst \citep{Yun94}.  The presence of extraplanar H~\textsc{i} surrounding  M82 is likely to also be a result of this interaction  \citep{Yun93}.

The stellar body of M82 is a high surface brightness disk with a moderate radial scale length that is only mildly distorted in its outer regions  \citep{Ichikawa95}.  The kinematics of the central region are dominated by the presence of a bar that extends over much of the $\sim$400~pc diameter starburst region \citep{Wills97,Greve02} and is surrounded by a molecular gas ring or tightly wound arms.  This region contains the majority of the molecular mass of $\sim 3 \times 10^{8}$ M$_{\odot}$ \citep{Naylor10}, which is comparable to the mass of H \textsc{i} surrounding the galaxy \citep{Yun93}.  The star-formation rate within the starburst region of $\sim$10~M$_{\odot}$~yr$^{-1}$ \citep[e.g.,][]{Gao04,Forster03b} exceeds that of the entire Milky Way.  M82 thus should exhaust its internal gas supply in $\sim$0.1~Gyr, which is significantly less than the time scales of typical galaxies.

M82's interstellar medium consists of compact, dense molecular clouds and dense warm, ionized gas, both embedded in a hot, low density medium in approximate overall pressure equilibrium \citep{West09}.  Pressures in the starburst region are extremely high, on the order of 0.5-1.0$\times 10^{7}$ K~cm$^{-3}$ \citep{O'Connell78,Smith06,West07}.  Observations of CO lines suggest molecular gas densities of 10$^{3}$ - 10$^{4}$ cm$^{-3}$ with kinetic temperatures of $\> 40$ K \citep{Wild92,Mao00}, whereas triatomic molecules such as HCO and HNC naturally suggest higher densities between $10^{4.0}$ and $10^{4.8}$ cm$^{-3}$ with temperatures between 50 and 500 K \citep[e.g.,][]{Muehle07,Fuente08,Naylor10}.  CO measurements also constrain the mass of the molecular gas to $3 \pm 1 \times 10^{8}$ M$_{\odot}$ \citep{Naylor10, Wild92}, much of which may be in the form of fragmented clouds.  Observations in the near-infrared give warm, ionized gas densities on the order of $\sim$10-600 cm$^{-3}$ with a gas mass of $\sim 8 \times 10^{6}$ M$_{\odot}$ \citep{Forster01,West09}.

Due its intense star formation, M82 produces its bolometric luminosity of $\sim 4 \times 10^{10}$ L$_{\odot}$ in only a small region. Assuming a cylindrical geometry with a radius of 200 pc and scale height of 100~pc, we find an internal radiation field energy density of U$_{rad} \approx 10^{3}$ eV~cm$^{-3}$ in the starburst zone.  Thus both the thermal pressure and energy density are substantially enhanced in M82 relative to these quantities in the ISM near the Sun.  If the magnetic field energy density also is increased by factors of $\sim 10^{3}$ over that in the solar neighborhood, then we would expect to find magnetic fields in M82 with average strengths of $B \sim 150 ~ \mu$G.


\subsection{Galactic Wind}

M82 is well known for its highly visible galactic wind, which has been observed from the x-rays to the radio \citep[e.g.,][]{O'Connell78,Seaquist91,Ohyama02,Stevens03,Engelbracht06,Strickland07}. At most wavelengths the outflow has an approximately bipolar structure that emerges nearly perpendicular to M82's stellar disk.  Optical observations of emission lines from ionized gas entrained in the wind indicate outflow speeds of $\sim$500-600 km~s$^{-1}$ \citep[][and references therein]{Shopbell98}.  Fits to the x-rays, however, indicate higher terminal wind speeds of up to $\sim$1400-2200 km~s$^{-1}$ \citep{Strickland09}.  Traditionally the M82 wind has been interpreted as a hot outflow powered by shock heating from supernovae \citep{Chevalier85}, but cosmic ray also could be a factor in driving the wind \citep{Breit93}.  The variation in speeds and excitation levels within the M82 wind indicate that it has a complex structure, and so may also be subject to a variety of driving mechanisms.


\subsection{Supernova Rate and Cosmic Ray Injection}

An advantage of M82 for cosmic ray studies is that the supernova rate can be directly determined through radio studies of the numbers and expansion rates of supernova remnants. Although values as high as $\nu_{SN} =$ 0.2-0.3 SN~yr$^{-1}$ have been used for the supernova rate in M82 \citep{delPozo09b}, we adopt the more conservative estimate of $\nu_{SN} =$ 0.07 yr$^{-1}$ from \cite{Fenech08}.  This is likely to be a lower bound on the true rate, and is consistent with other estimates of the M82 star formation rate \citep{Fenech10}.  

Supernovae are the assumed accelerators of cosmic rays.  Of the energy resulting from the explosion, only a small fraction is transferred to cosmic rays.  In the Milky Way, the typical value for this efficiency factor, $\eta$, is 10\% \citep{Blandford87}.  We discuss models for $\eta$ = 0.04-0.2 in \S 5.  The majority of the energy transferred to cosmic rays goes into protons and only a small fraction goes into electrons.  The ratio of protons to electrons is here assumed to be $N_{p}/N_{e} \sim 50$ \citep{Blandford87}.  The spectrum of cosmic rays accelerated by supernovae is generally assumed to be a power law with spectral index $p \sim 2.1-2.2$.


\subsection{Radio Continuum}

Of the available radio data sets, a collection of single-dish observations by \cite{Klein88} has the largest range of frequencies (from $10^{7}$ to $10^{14}$ Hz).  Thermal dust emission becomes important starting at $\sim$$10^{11}$ Hz.  The lowest frequency data will likely overestimate the flux of the starburst due to a large beam size.  Because of this we only consider data from $10^{8}$ to $10^{11}$ Hz in the Klein et al. data set.  More recently, \cite{Williams10} observed the starburst region from 1 to 7~GHz with the Allen Telescope Array.  These interferometer measurements have significantly smaller fluxes than the single-dish observations: the interferometer fluxes are $\sim$12\% smaller (see Section 4).


\subsection{Gamma Rays}

Starburst galaxies were anticipated to be $\gamma$-ray sources because of their high star-formation and supernova rates.  Recently, $\gamma$-rays above 700 GeV were detected from M82 with the ground-based Cherenkov telescope VERITAS.  Measurements were made between $\sim$0.9 and $\sim$5~TeV with a flux upper limit (99\% confidence level) at $\sim$6.6~TeV \citep{Acciari09}.  M82 has also been detected in $\gamma$-rays with the \textit{Fermi $\gamma$-Ray Space Telescope} at energies from 200~MeV to 300~GeV \citep{Abdo10}.

\begin{figure*}[t!]
\epsscale{1.15}
\plottwo{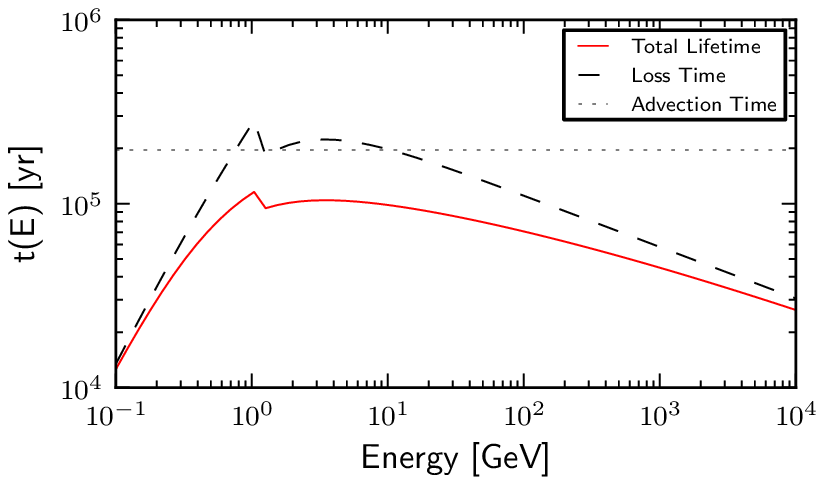}{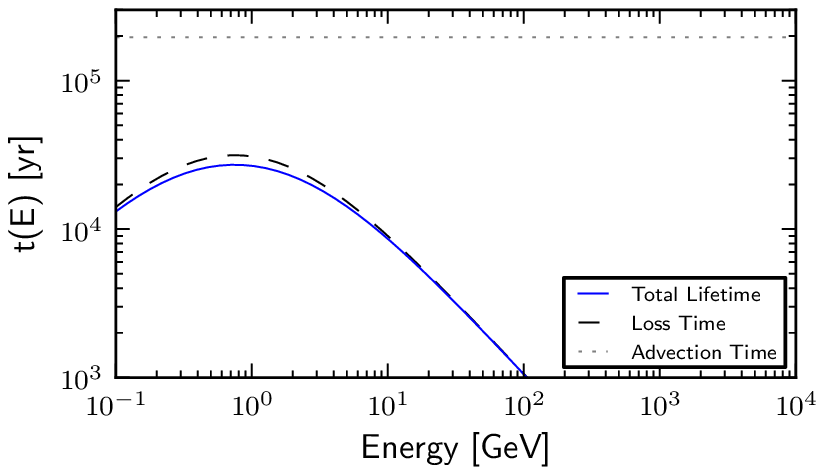}
\caption{Lifetimes for cosmic-ray protons (\textit{left}) and electrons (\textit{right}).  Dotted lines represent the non-interactive loss (advection) time for a wind speed of $v_{adv}$ = 500~km~s$^{-1}$.  Dashed lines represent the energy loss time.  Solid lines represent the combined particle lifetime.  The average ISM density is fixed at $n$ = 550~cm$^{-3}$ corresponding to a molecular gas mass of $4 \times 10^{8}$ M$_{\odot}$ (see Table 1 for additional parameters).  The discontinuity in the proton lifetime is due to the sudden turn on of pion production as an energy loss at 1.22 GeV.}
\end{figure*}
%
%


\section{Theoretical Approach}

To construct a model that predicts both the radio and $\gamma$-ray spectra, we must first calculate the cosmic ray spectra for M82.  We begin by assuming a source function for cosmic rays accelerated by supernovae.  We use a steady-state model and assume that diffusion will be relatively unimportant in the starburst region\footnote{Cosmic ray diffusion times in the Milky Way are about two orders of magnitude longer than the advection and loss times typical of our models for M82.}.  Then, we calculate the primary cosmic ray spectra by including a variety of energy losses and an advective galactic wind.  Because of the very dense nature of the interstellar medium in the starburst region and the relatively uniform distribution of supernovae in M82, we expect a high interaction rate for cosmic-ray protons.  As the dominant energy loss mechanism is pion production, we calculate the spectra for the secondary pions and their associated decay products (electrons, positrons, and $\gamma$-rays).  From these primary and secondary cosmic rays, we then predict the radio and $\gamma$-ray spectra.


\subsection{Primary Cosmic Rays}

The cosmic ray transport equation in a homogeneous medium \citep{Longair11}, with diffusion omitted and advective losses added, is
\begin{equation}\label{transport}
\frac{\partial N(E,t)}{\partial t} = - \frac{\partial}{\partial E} \left[ \frac{dE}{dt} N(E,t) \right] + Q(E,t)-\frac{N(E,t)}{\tau_{adv}},
\end{equation}
where $N(E,t) dE$ is the number density of particles with energy between $E$ and $E + dE$ at time $t$, $dE/dt < 0$ is the rate at which a particle's energy changes due to radiation and collisions, $Q(E,t)dE$ is the rate at which cosmic rays with energies between $E$ and $E + dE$ are injected per unit volume, and $\tau_{adv}$, which we assume is independent of $E$, is the rate at which particles are advected out of the region.

We assume $Q$ is independent of $t$ and seek time independent solutions of eqn. (\ref{transport}). Although the time independent version of eqn. (\ref{transport}) is a linear ODE, with  solutions that can be written down in closed form, the solutions are not very useful for our parameter study because the complicated form of $dE/dt$ makes them difficult to evaluate. Therefore, we make the approximation
\begin{equation}\label{N}
N(E) \approx Q(E) ~ \tau(E),
\end{equation}
where
\begin{equation}\label{tau}
\tau(E)^{-1} \equiv \tau_{adv}^{-1} + \tau_{loss}^{-1}
\end{equation}
is the total energy loss rate and
\begin{equation}\label{tauloss}
\tau_{loss} \equiv - \frac{E}{dE/dt}
\end{equation}
is the energy loss rate due to radiative and collisional processes. Equation (\ref{N}) becomes exact for $\tau_{adv} / \tau_{loss} \ll 1$, and is accurate to better than 10\% - 20\% in cases where $Q$ is a power law in $E$ with index near 2, as is thought to be the case for acceleration of cosmic rays by strong shocks, the leading theory of cosmic ray origin.

As supernovae are the assumed drivers of cosmic ray acceleration, the source function for cosmic rays must be related to the total energy input from supernovae.
\begin{equation}
\int_{E_{min}}^{E_{max}} Q(E) E dE = \frac{\eta \nu_{SN} E_{51}}{V} ,
\end{equation}
where $\nu_{SN}$ is the supernova rate, $V$ is the volume of the starburst region, $\eta$ is the fraction of the supernova energy transferred to cosmic rays, and $E_{51} = 1$ is 10$^{51}$ ergs, the typical energy from a supernova explosion.  Then, assuming the source function is a power law of the form $Q(E) \propto E^{-p}$, the particle spectrum becomes
\begin{equation}
N(E) = \frac{(p-2)}{E_{min}^{-p+2}} ~ \frac{\eta \nu_{SN} E_{51}}{V} E^{-p} ~ \tau(E).
\end{equation}
Energy losses for protons include ionization, Coulomb interactions, and pion production.  For electrons, losses include ionization, bremsstrahlung, inverse Compton scattering, and synchrotron radiation.  Cosmic ray lifetimes for values representative of our best fit models are plotted in Figure 1, and energy loss rates are plotted in Figure 2. Energy loss rates are explained in the Appendix.

\begin{figure*}[t!]
\epsscale{1.15}
\plottwo{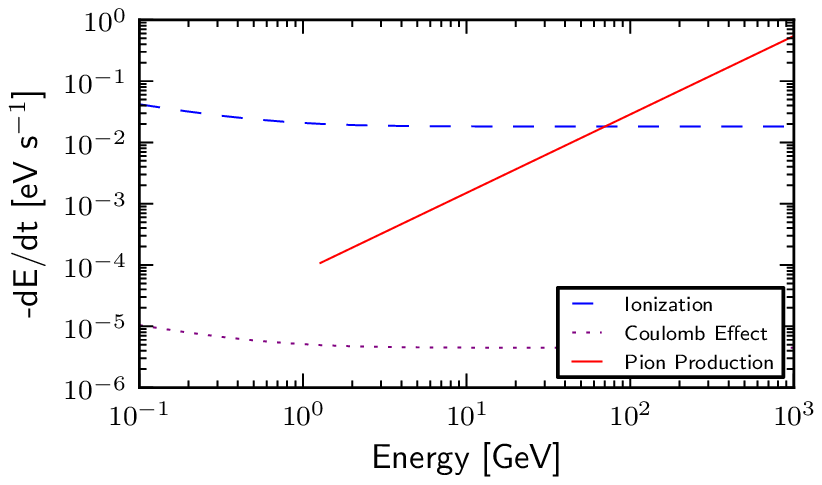}{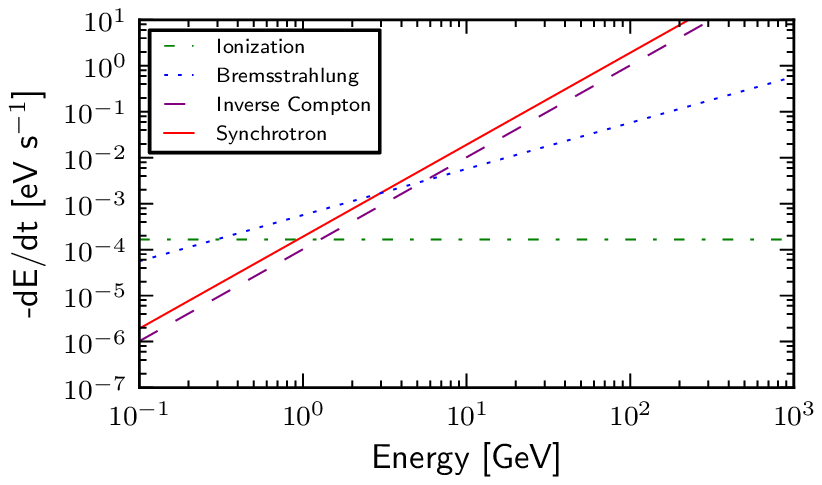}
\caption{Rate of energy loss for protons (\textit{left}) and electrons (\textit{right}).  Proton losses include ionization (dashed blue line), the Coulomb effect (dotted purple line), and pion production (solid red line).  Electron losses include ionization (dot-dashed green line), bremsstrahlung (dotted blue line), inverse Compton scattering (dashed purple line), and synchrotron emission (solid red line).  A magnetic field strength of  $B$ = 275~$\mu$G was selected for this example (see \S 4.4) with the fixed average density of $n$ = 550~cm$^{-3}$ and a radiation field energy density of $U_{rad} = 1000$ eV~cm$^{-3}$.}
\end{figure*}

Only a small fraction of the supernova power goes into cosmic-ray electrons, as compared to cosmic-ray protons.  We assume the source functions for protons and electrons can be related by \citep{Torres04}
\begin{equation}
Q_{e}(E) = \frac{N_{e}}{N_{p}} Q_{p}(E).
\end{equation}
We assume the ratio of protons to electrons to be $N_{p} / N_{e} = 50$.


\subsection{Secondaries}

In this section, we discuss pions and the particles resulting from their decay.  Above $\sim 1$ GeV, the dominant energy loss for cosmic-ray protons is pion production, and so in this regime, the source function for pions can be calculated from the proton spectrum.  Assuming a delta function approximation, which is appropriate for smooth power-law distributions of cosmic rays, the source function for pions is given by \citep{Kelner06}
\begin{equation}
q_{\pi}(E_{\pi}) = c n \int \delta(E_{\pi} - K_{\pi}T_{p}) \sigma_{pp}(E_{p}) N_{p}(E_{p}) dE_{p}.
\end{equation}
Integrating over proton energy, this becomes
\begin{equation}
q_{\pi} (E_{\pi}) = \tilde{n} \frac{c n}{K_{\pi}} \sigma_{pp} \left(m_{p}c^{2} + \frac{E_{\pi}}{K_{\pi}} \right) N_{p} \left(m_{p}c^{2} + \frac{E_{\pi}}{K_{\pi}} \right),
\end{equation}
where $K_{\pi} \approx 0.17$ is the mean fraction of the kinetic energy of the proton, $T_{p}$, transferred to the pion per collision.

While the cross sections for pion production from proton-proton interactions have been widely studied, there is no clear agreement as to which parameterization should be used.  Of those available, we considered the model of \cite{Kamae05} which includes the effects of the diffraction dissociation process and scaling violations, and the model of \cite{Kelner06}, whose energy weighted total cross section is based on the numerical simulations of proton-proton interactions by the \textsc{sibyll} code.  Comparing the two cross sections, the models disagree at energies between 1 and 10 GeV.  However, the $\gamma$-ray source functions calculated from the two models are nearly identical.  The differences in cross section add less than a factor of $\sim$2 in uncertainty \citep{Domingo05, Lacki13}.  As the \citet{Kelner06} and \citet{Kamae05} cross sections give such similar results, we chose to use \citeauthor{Kelner06}'s model for simplicity.  

\cite{Kelner06} give the total inelastic cross section as
\begin{align}
\sigma_{inel} (E_{p}) ~ = ~ &(34.3 ~ + ~ 1.88L ~ + ~ 0.25L^{2} ) \nonumber \\ &\times \left[ 1 - \left( \frac{E_{th}}{E_{p}} \right)^{4} \right]^{2} ~ \text{mb},
\end{align}
where  $L$ = $ln(E_{p}$/TeV).  This gives us the cross section for neutral pion production.  For charged pions, we can multiply the inelastic cross section by the multiplicity of the charged pions.  To obtain the multiplicity, we take the ratio of the inclusive cross sections for the production of charged pions to the inclusive cross section for the production of neutral pions.  We use the inclusive cross sections modeled in \cite{Dermer86} for this purpose.

Charged pions are relatively short lived and quickly decay into muons and then into electrons, positrons, and neutrinos, $\pi^{+} \rightarrow \mu^{+} + \nu_{\mu}$ and $\mu^{+} \rightarrow e^{+} + \nu_{e} + \overline{\nu}_{\mu}$.  Because the muon moves with nearly the same speed as the pion, in the laboratory frame, their source functions are equivalent, $q_{\mu}(\gamma_{\mu}) \simeq q_{\pi}(\gamma_{\pi})$.  Then, the secondary electron and positron source functions are given by
\begin{equation}
q_{e}(\gamma_{e}) = \int_{1}^{\gamma_{e}'(max)} d\gamma_{e}' \frac{P(\gamma_{e}')}{2\sqrt{\gamma_{e}'^{2}-1}} \int_{\gamma_{\mu}^{-}}^{\gamma_{\mu}^{+}} d\gamma_{\mu} \frac{q_{\mu}(\gamma_{\mu})}{\sqrt{\gamma_{\mu}^{2}-1}},
\end{equation}
where $\gamma_{\mu}^{\pm} = \gamma_{e}\gamma_{e}' \pm \sqrt{\gamma_{e}^{2}-1} \sqrt{\gamma_{e}'^{2}-1}$, $\gamma_{e}'(max) = 104$ and the electron/positron distribution in the muon's rest frame is (see \citealt{Schlick02})
\begin{equation*}
P(\gamma_{e}') = \frac{2 \gamma_{e}'^{2}}{\gamma_{e}'^{3}(max)} \left( 3 - \frac{2 \gamma_{e}'}{\gamma_{e}'(max)} \right).
\end{equation*}
Primary and secondary electron and positron spectra can be seen in Figure 3.

Neutral pions have extremely short lifetimes and quickly decay into $\gamma$-rays, $\pi^{0} \rightarrow \gamma + \gamma$.  The source function for $\gamma$-rays is given by \citep{Schlick02,Pohl94}
\begin{equation}
q_{\gamma,\pi}(E_{\gamma}) = 2 \int_{E_{min}}^{\infty} \frac{q_{\pi^{0}}(E_{\pi})}{\sqrt{E_{\pi}^{2} - m_{\pi}^{2}c^{4}}} dE_{\pi},
\end{equation}
where $E_{min} = E_{\gamma} + (m_{\pi}c^{2})^{2}/(4 E_{\gamma})$.

\begin{figure*}[ht!]
\epsscale{1.15}
\plottwo{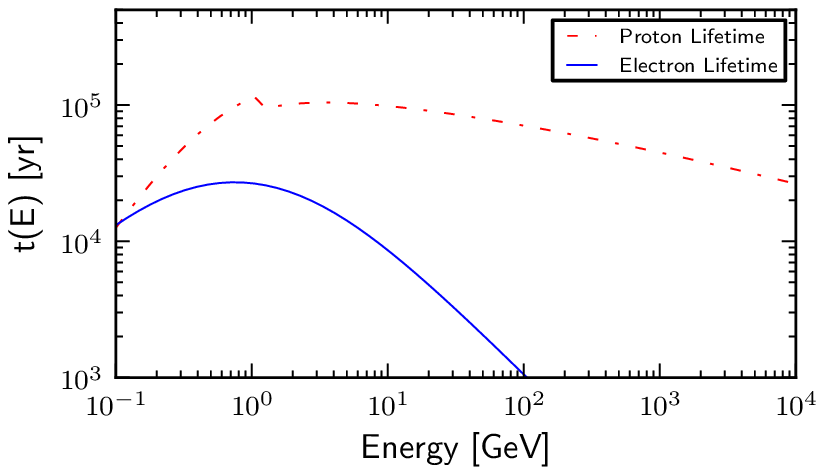}{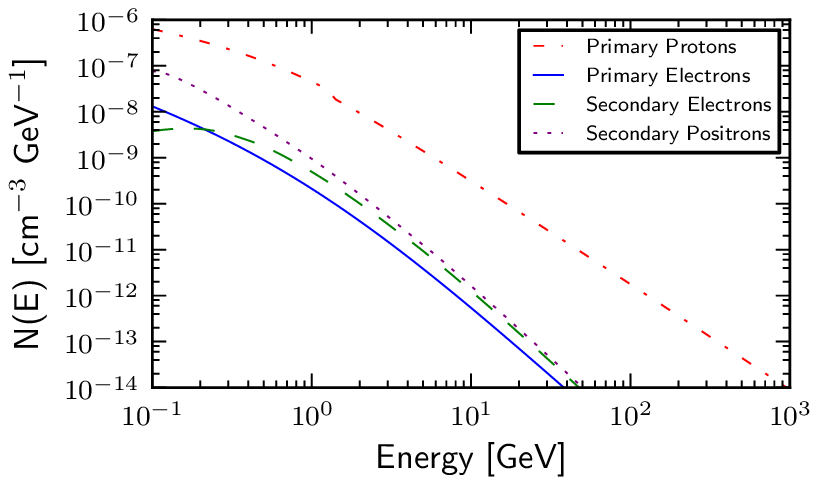}
\caption{\textit{Left:} Total lifetimes for cosmic-ray protons and electrons.  \textit{Right:} Spectra for primary cosmic-ray protons and electrons and secondary cosmic-ray electrons and positrons.  Dot-dashed red lines represent protons, solid blue lines represent primary electrons, dashed green lines represent secondary electrons, dotted purple lines represent secondary positrons.  Parameters were set at $B$ = 275~$\mu$G, $v_{adv}$ = 500~km~s$^{-1}$, with $n$ = 550~cm$^{-3}$ and $p$ = 2.1.}
\end{figure*}
%
%


\subsection{$\gamma$-Rays}

In addition to production by the decay of neutral pions, $\gamma$-rays are also produced by inverse Compton scattering and by relativistic, non-thermal bremsstrahlung from cosmic-ray electrons.  In the case of production by bremsstrahlung, the source function for the $\gamma$-rays is given by \citep{Stecker71}
\begin{equation}
q_{\gamma,brem}(E_{\gamma}) = c n \sigma_{brem} E_{\gamma}^{-1} \int_{E_{\gamma}}^{\infty} N_{e}(E_{e}) dE_{e},
\end{equation}
where $\sigma_{brem} = 3.38 \times 10^{-26}$ cm$^{2}$ and $N_{e}(E_{e})$ represents the combined cosmic-ray electron/positron spectrum.

For inverse Compton scattering, the $\gamma$-ray source function is given by \citep{RL79}
\begin{equation}
q_{\gamma,IC}(E_{\gamma}) = \frac{3 c \sigma_{T}}{16 \pi} \int_{0}^{\infty} d\epsilon \frac{v(\epsilon)}{\epsilon} \int_{\gamma_{min}}^{\infty} d\gamma \frac{n_{e}(\gamma)}{\gamma^{2}} F(q, \Gamma)
\end{equation}
with \citep{Schlick02}
\begin{equation*}
\gamma_{min} = \frac{E_{\gamma}}{(2 m_{e} c^{2})} \left[ 1 + \left( 1 + \frac{m_{e}^{2}c^{4}}{\epsilon E_{\gamma}} \right)^{1/2} \right],
\end{equation*}
where $E_{\gamma}$ is the energy of the resulting $\gamma$-ray, $\epsilon$ is the energy of the incident photon, $\gamma$ is the energy of the electron.  Here, the electron spectrum, $n_{e}(\gamma)$, is in units of cm$^{-3}$.  The function $F(q, \Gamma)$ is part of the Klein-Nishina cross section and is given by \citep{Blumenthal70}
\begin{equation*}
F(q, \Gamma) = 2q \text{ln}(q) + (1 + q - 2q^{2}) + \frac{\Gamma^{2}q^{2} (1 - q)}{2(1 + \Gamma q)},
\end{equation*}
where
\begin{equation*}
\Gamma = \frac{4 \epsilon \gamma}{(m_{e}c^{2})} \text{ and } q = \frac{E_{\gamma}}{\Gamma (\gamma m_{e}c^{2} - E_{\gamma})}.  
\end{equation*}
For the blackbody spectrum, $v(\epsilon)$, we use an isotropic, diluted, modified blackbody spectrum \citep[][and references therein]{Persic08}
\begin{equation}
v(\epsilon) = \frac{C_{dil}}{\pi^{2} \hbar^{3} c^{3}} \frac{\epsilon^{2}}{e^{\epsilon / k T_{d}} - 1} \left( \frac{\epsilon}{\epsilon_{0}} \right)^{\sigma = 1},
\end{equation}
where $C_{dil}$ is a spatial dilution factor (given by the normalization $U_{rad} = \int v(\epsilon) \epsilon d\epsilon$) and $\epsilon_{0}$ corresponds to $\nu = 2 \times 10^{12}$ Hz.

\begin{figure*}[t!]
\epsscale{0.9}
\plotone{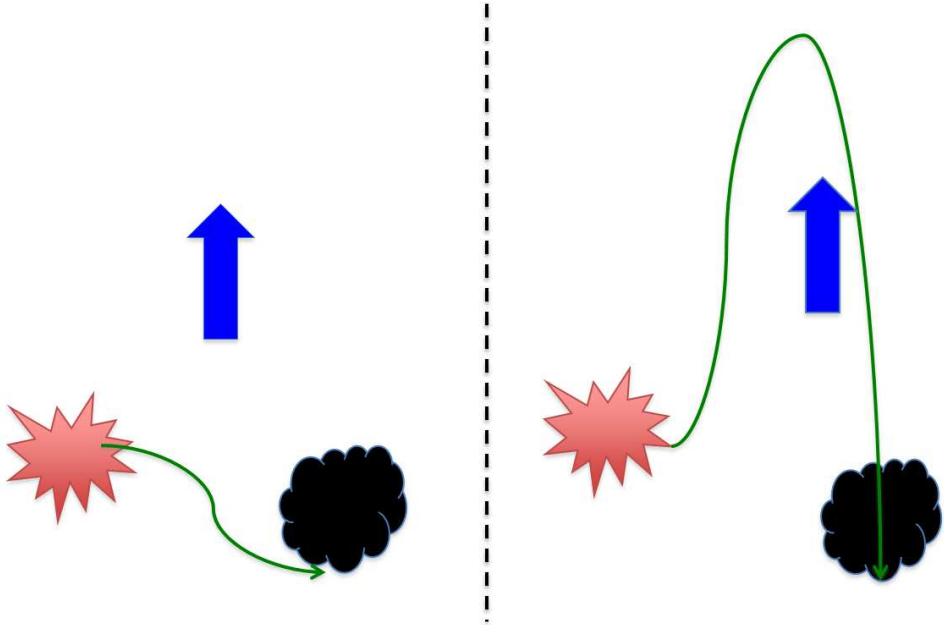}
\caption{Schematic of the process. The red explosion is a supernova remnant, the black cloud is a molecular cloud, the green line is a schematic magnetic fieldline loaded with cosmic rays, and the blue arrow represents a galactic wind flow. The figure on the left shows an initial configuration, and the figure on the right shows the configuration after the fieldline has been carried out some distance by the wind.}
\end{figure*}
%
%


\subsection{Radio Spectrum}

We expect that the majority of the radio flux comes from non-thermal synchrotron emission and thermal free-free emission.  To calculate the radio synchrotron spectrum, we utilize the emission coefficient given by \cite{Longair11},
\begin{equation}
j_{\nu}^{synch} = \left(- \frac{dE}{dt} \right) N(E) \frac{dE}{d\nu},
\end{equation}
where N(E) is the combined electron and positron spectrum and
\begin{equation}
- \frac{dE}{dt} = \frac{4}{3} \sigma_{T} c \left( \frac{E}{m_{e}c^{2}} \right)^{2} \frac{B^{2}}{2 \mu_{0}}
\end{equation}
gives the synchrotron losses as a function of energy.  If we make the simplifying assumption that an electron of energy $E$ radiates away its energy at the critical frequency, $\nu_{c}$, such that $\nu \approx \nu_{c} \approx \gamma^{2} \nu_{g} \approx \gamma^{2} \cdot eB/ (2 \pi m_{e})$, then
\begin{equation}
j_{\nu}^{synch} = \frac{4}{3} \sigma_{T} m_{e} c^{3} \left( \frac{\nu}{\nu_{g}^{3}} \right)^{1/2} \frac{B^{2}}{2 \mu_{0}} N(E).
\end{equation}
Because of the warm, ionized gas in the ISM, we expect some portion of the radio spectrum to come from thermal bremsstrahlung or free-free emission and absorption.  The free-free emission coefficient is given by \citep{RL79}
\begin{equation}
j_{\nu}^{ff} = 6.8 \times 10^{-38} T^{-1/2} Z^{2} n_{e} n_{i} e^{-h\nu / k_{B}T} \overline{g}_{ff}
\end{equation}
with units of ergs~s$^{-1}$~cm$^{-3}$~Hz$^{-1}$.  Additionally, we must consider the inverse process, free-free absorption.  The absorption coefficient is given by \citep{RL79}
\begin{equation}
\kappa_{\nu}^{ff} = 0.018 T^{-3/2} Z^{2} n_{e} n_{i} \nu^{-2} \overline{g}_{ff} ~~ \text{cm}^{-1}.
\end{equation}
In the ``small-angle, classical region," the mean Gaunt factor is \citep{Novikov73}
\begin{equation}
\overline{g}_{ff} = \frac{\sqrt{3}}{\pi} ln \left[ \frac{1}{4 \zeta^{5/2}} \left( \frac{k_{B}T}{h \nu} \right) \left( \frac{k_{B}T}{Z^{2} Ry} \right)^{1/2} \right],
\end{equation}
where $Ry = 13.6$ eV is the Rydberg constant and $\zeta = e^{\gamma} \approx 1.781$ with $\gamma$ being Euler's constant.

For both emission and absorption, the radiative transfer equation is \citep{RL79}
\begin{equation}\label{radtrans}
\frac{dI_{\nu}}{d\tau_{\nu}} = -I_{\nu} + S_{\nu},
\end{equation}
where $S_{\nu} \equiv j_{\nu}/\kappa_{\nu}$ is the source function.  We discuss solutions to eqn. (\ref{radtrans}) in \S 4.3.1.


\section{Model and Results}

We have calculated the steady-state primary and secondary cosmic ray spectra according to the assumptions made in Section 3.  From there, we compared the observed properties of M82 with our model.  We included the multiphase nature of the interstellar medium in order to better predict absorption and emission in the radio spectrum.  We calculated the synchrotron emission from the primary and secondary electrons and positrons and we also calculated the effects of free-free emission and absorption on the radio spectrum.  In this section, for both the radio and $\gamma$-ray spectra, we discuss the effects and constraints of various parameters on the predicted spectra and the resulting best-fit models.


\subsection{Cosmic Ray Propagation in the Disk}

We assume that the cosmic rays sample the mean density in the disk.  This assumption is not trivial, since we expect that most of the cosmic rays are accelerated in the hot, low density medium, which is flowing out of the disk, while most of the mass is in the form of dense, molecular gas which has a small filling factor.  Cosmic rays will sample the mean density if (1) most of the magnetic field lines they are accelerated on connect to a ``fair sample" of the interstellar gas, and (2) cosmic rays can propagate a sufficient distance along the fieldlines, or diffuse sufficiently across the fieldlines, to encounter dense gas  before they - and the fieldlines -  are borne out of the disk by the wind (see Figure 4).  The timescale for the latter is $\tau_{adv}$, the escape time introduced in eqn.~(5).

By a ``fair sample," we mean that the ratio of interstellar gas mass to magnetic flux is uniform within the starburst region.  Since the covering factor of the molecular clouds is below unity, this suggests that the fieldlines have a substantial random component, as is the case in the Milky Way.  In the absence of any information to the contrary, we assume the fieldlines are indeed uniformly loaded, so that the average gas density on any magnetic flux tube within the disk is the mean interstellar density, $f_{mol} n_{mol} + f_{ion} n_{ion} + (1-f_{mol}-f_{ion}) n_{hot} \approx f_{mol} n_{mol}$.  If there are $N_{mc}$ molecular clouds in the disk and the mean dimension of the disk is $L \sim V^{1/3}$ where $V = 2 \pi R^{2} H$, then the mean distance $l_c$ to a cloud is of order $L N_{mc}^{-1/3}$.  If the supernova remnants where the cosmic rays are accelerated and the molecular clouds are spatially uncorrelated, then $l_c$ is the mean minimum distance that cosmic rays must  travel to encounter dense gas.

\begin{figure*}[ht!]
\epsscale{1.15}
\plottwo{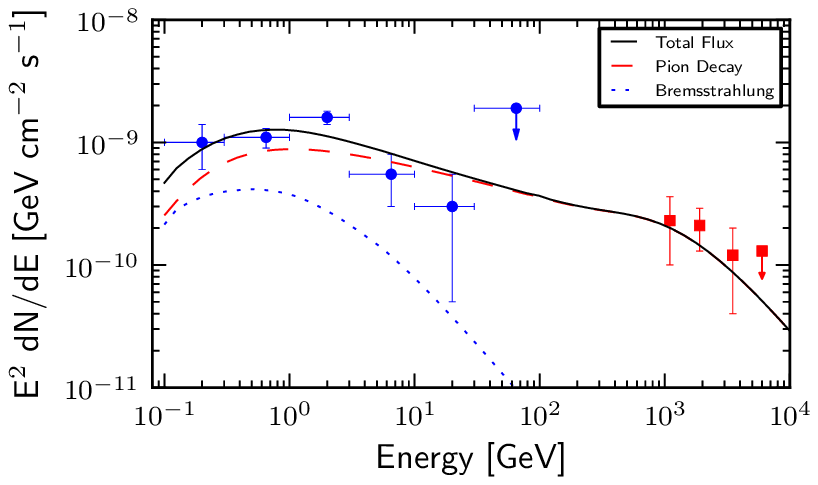}{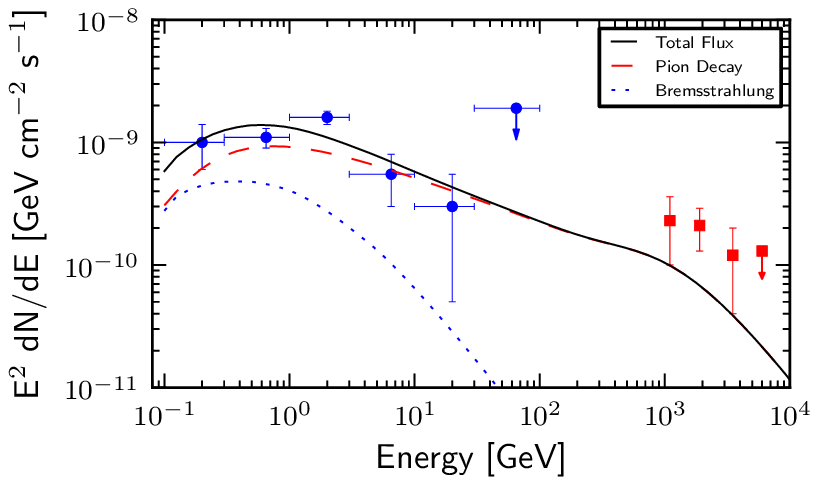}
\caption{$\gamma$-ray spectra. \textit{Left:} $\gamma$-ray spectrum with parameters $p$ = 2.1, $M_{mol} = 4 \times 10^{8}$~M$_{\odot}$, $B$ = 275~$\mu$G, $v_{adv}$ = 500~km~s$^{-1}$, $n_{ion}$ = 100~cm$^{-3}$.  This spectrum is the best fit to the $\gamma$-ray data.  \textit{Right:} $\gamma$-ray spectrum with parameters $p$ = 2.2, $M_{mol} = 4 \times 10^{8}$~M$_{\odot}$, $B$ = 275~$\mu$G, $v_{adv}$ = 400~km~s$^{-1}$, $n_{ion}$ = 150~cm$^{-3}$.  This is the $\gamma$-ray spectrum that corresponds to the best radio fit for a spectral index of $p$ = 2.2.  The solid black lines represent the total $\gamma$-ray flux, the dashed red lines represent the contribution from neutral pion decay, and the dotted blue lines represent the contribution from bremsstrahlung.  $\gamma$-ray data include: \cite{Ackermann12} (\textit{Fermi} - blue circles), \cite{Acciari09} (VERITAS - red squares).  Data with downward arrows represent upper limits for both Fermi and VERITAS data.}
\end{figure*}

In order to estimate the time it takes cosmic ray particles to travel a distance $l_c$, we assume the cosmic rays are self-confined (see \cite{Kulsrud05} for a pedagogical treatment): any cosmic ray density gradient produced by a point source of cosmic rays, or by the global distribution of cosmic ray sources within the starburst region, is associated with an anisotropy of the cosmic rays in velocity space.  If the mean cosmic ray velocity, or streaming speed, exceeds the Alfven speed $v_A$, then the cosmic rays destabilize short wavelength Alfven waves.  The waves grow large enough to scatter the cosmic rays, reducing their drift speed to $v_A$.  In this picture, the propagation time is $l_c/v_A$.  The cosmic rays will encounter dense clouds if
\begin{equation}\label{criterion}
\frac{v_A \tau_{adv}}{l_c} \sim \frac{v_A}{v_{adv}} \left( \frac{H}{R} \right)^{2/3} \frac{N_{mc}^{1/3}}{(2 \pi)^{1/3}} > 1,
\end{equation}
where in the last step we have identified $\tau_{adv}$ with an advection time $H/v_{adv}$.

In \S 4.4 we will show that both the radio and $\gamma$-ray spectra are very well fit by a simple family of models in which the magnetic field strength $B$ and advection velocity $v_{adv}$ are related by the empirical formula
\begin{equation}\label{bestfit}
v_{adv} = 4.0 B_{\mu} - 600,
\end{equation}
where $B_{\mu}$ is the magnetic field in $\mu$G and $v_{adv}$ is in km~s$^{-1}$. Equation~(\ref{bestfit}) holds for $150 < B_{\mu} < 350$.  Substituting this relation into eqn.~(\ref{criterion}), using $H/R = 0.5$ (see Table 1) and solving for $N_{mc}$, we find that cosmic rays will encounter dense clouds before being advected out of the Galaxy if
\begin{equation}\label{criterion2}
N_{mc} > 8 \pi \left( 2.00 - \frac{300}{B_{\mu}} \right)^3 n_{hot}^{3/2}.
\end{equation}
According to eqn.~(\ref{criterion2}), for the value of $N_{mc}$ needed to satisfy the criterion varies from 0 to 7 as $B_{\mu}$ ranges from 150 to 350 with $n_{hot} = 0.33$ cm$^{-3}$ (see Table 1).  With the molecular gas mass of the M82 starburst region estimated to be about $3 \times 10^{8}$~M$_{\odot}$, and giant molecular cloud masses in the Milky Way estimated at 10$^{4 - 6}$~M$_{\odot}$, this criterion would seem to be easily satisfied.  Moreover, the drift speed actually exceeds $v_A$ if the waves are strongly damped, making eqn.~(\ref{criterion}) even easier to satisfy.  A more careful assessment based on a Monte Carlo simulation supports this conclusion \citep{Boettcher}.  This effect may be important for the higher energy cosmic rays.  Once the cosmic rays reach the weakly ionized clouds, the waves which scatter them are strongly damped, and the cosmic rays interact primarily through physical collisions \citep{Everett11}.


\subsection{$\gamma$-Rays}

Though we incorporate the multiphase nature of the interstellar medium into our model for the radio spectrum (see \S 4.3.1), this is not necessary in the case of the $\gamma$-rays because of our assumption that they sample the mean density (\S 4.1) which is almost entirely due to molecular clouds.  The ionized gas mass is well-constrained by observations to be $M_{ion} = 8 \times 10^{6}$ M$_{\odot}$.  In \S 4.4 we describe models with molecular masses of $M_{mol} = 2-4 \times 10^{8}$ M$_{\odot}$ corresponding to mean densities of $\langle n \rangle = 280-550$ cm$^{-3}$.

Pion decay is the main mechanism for the production of $\gamma$-rays.  When advective losses dominate over energy losses, the $\gamma$-ray flux is proportional to the gas density.  However, in the calorimeter limit, where cosmic-ray proton energy losses dominate over advective losses, the $\gamma$-ray flux is essentially independent of density and so the number of $\gamma$-rays depends only on the number of protons.  Other parameters that affect the $\gamma$-ray spectrum include the supernova rate ($\nu_{SN}$ = 0.07-0.1 SN~yr$^{-1}$) and the distance ($d$ = 3.5-3.9 Mpc), both of which are observationally constrained.

In addition to the supernova rate and distance, wind advection speed ($v_{adv}$) affects the $\gamma$-ray spectrum.  For our cosmic ray spectra, we use a combined lifetime (see eqn. (\ref{tau})) which includes energy losses and advection.  The advection time is given by $\tau_{adv} = H / v_{adv}$ where H is the height of the starburst region and $v_{adv}$ is the speed of particles in the wind in the starburst region.  Of the parameters that affect the $\gamma$-ray spectrum, the least constrained is the wind speed.  For the $\gamma$-ray spectrum, the main constraint on the wind speed is the Fermi observations.  The proton lifetime peaks between 1 to 10 GeV, and it is this portion of the proton spectrum that contributes to the energy range for the $\gamma$-ray spectrum observed by Fermi (0.3 to 30 GeV).  So, if the wind speed becomes too large, the maximum proton lifetime decreases and the $\gamma$-ray spectrum will no longer agree with the Fermi observations.  If this is the case, then the galaxy is no longer calorimetric.

Assuming values typical of the Milky Way for parameters related to cosmic ray acceleration by supernovae, we find that the $\gamma$-ray flux from neutral pion decay fits the TeV energy $\gamma$-ray data and all but the two lowest-energy Fermi data points as seen in Figure 5 and discussed in \S 4.4.  As the $\gamma$-ray flux from neutral pion decay does not account for all of the observed flux below $\sim$1 GeV, there must be another emission mechanism responsible for the total $\gamma$-ray flux at these energies.  Two possible mechanisms are inverse Compton emission and bremsstrahlung.  For $\gamma$-ray emission from the inverse Compton effect, we found that the $\gamma$-ray flux in the starburst region was on the order of $2 \times 10^{-10}$ GeV~cm$^{-2}$~s$^{-1}$ at $\sim$1~GeV assuming an approximate radiation field energy density of 1000 eV~cm$^{-3}$ (see Table 1).  While our model currently includes inverse Compton emission from only primary electrons, we found that it to be a significant contribution from the total $\gamma$-ray flux at both GeV and TeV energies.  Inclusion of emission from secondary electrons and positrons will increase the total $\gamma$-ray flux by a factor of ~3 at minimum.  We will provide a more accurate model for inverse Compton emission in future work.

We also calculated the $\gamma$-ray flux from non-thermal, relativistic bremsstrahlung.  As bremsstrahlung depends on the electron spectrum, the flux is dependent upon the magnetic field strength and wind speed.  We find that bremsstrahlung makes a significant contribution to the $\gamma$-ray flux below $\sim$10~GeV (see Figure 5).  Thus, to accurately model the observed $\gamma$-ray flux at lower energies, we must include both neutral pion decay and bremsstrahlung in our work.


\subsection{Radio Spectrum}

In the case of the radio spectrum, we have several more parameters that affect our fit than for the $\gamma$-ray spectrum.  These parameters are the ionized gas density ($n_{ion}$), the magnetic field strength ($B$), and the wind speed ($v_{adv}$).  Free-free emission and absorption both scale as the square of the gas density.  Thus, the lower the gas density, the lower the frequency at which the radio spectrum turns over and the lower the contribution of free-free emission to the radio spectrum at higher frequencies.  Additionally, synchrotron emission is scaled by the square of the magnetic field strength ($B$), and the wind speed ($v_{adv}$) affects the cosmic ray ``dwell time" in the starburst.  To quantitatively distinguish between the effects of our parameters, we use $\chi^{2}$ tests to calculate the goodness-of-fit.  

\begin{figure*}[t!]
\epsscale{0.9}
\plotone{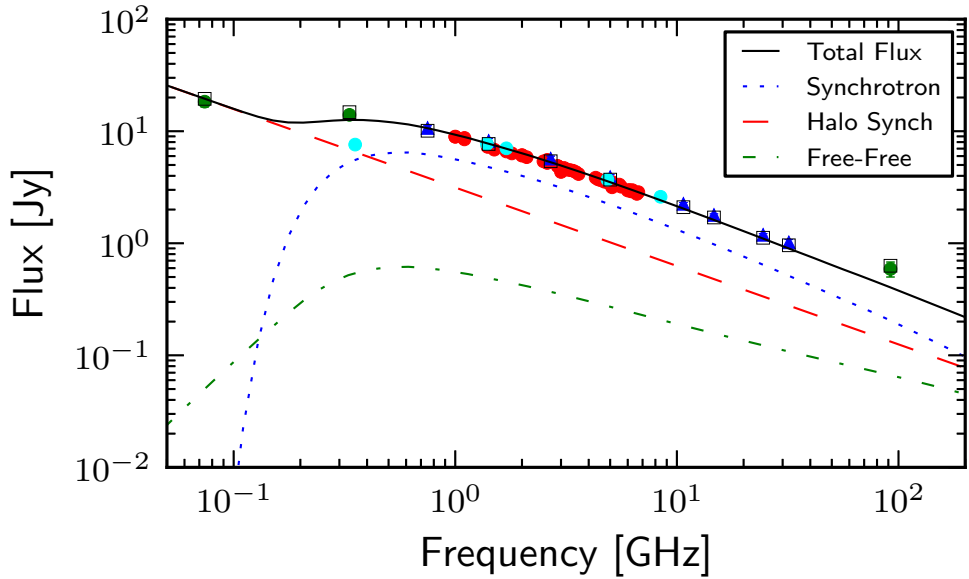}
\caption{Best fit to radio spectrum ($\chi^{2} = 22.6$) with parameters $B$ = 275~$\mu$G, $v_{adv}$ = 500~km~s$^{-1}$, $n_{ion}$ = 100~cm$^{-3}$ with $p$ = 2.1, $M_{mol}$ = $4 \times 10^{8}$~M$_{\odot}$.  The solid black line denotes total radio flux, the red dashed line represents the radio emission in the halo, the dotted blue line represents the radio emission in the hot, diffuse gas, and the dot-dashed green line represents radio emission in the warm, ionized gas.  Radio data include \cite{Klein88} (blue triangles), \cite{Williams10} (red circles), \cite{Cohen07} (green circle), \cite{Basu12} (green circle), \cite{Carlstrom91} (green circle), \cite{Adebahr} (cyan circle).  Triangles represent single-dish data and circles represent interferometer data.  Open squares represent the median data set used for the $\chi^{2}$ tests.}
\end{figure*}
%
%

\subsubsection{Multiphase Solutions for Radiative Transfer}

The emergent radio spectrum is due to free-free and synchrotron emission, modified by free-free absorption as discussed in \S 3.3.  Fitting a model to the observed radio spectral energy distribution therefore depends on knowing the degree of free-free absorption.  This is most pronounced at lower frequencies and these observations are essential for obtaining a physically realistic fit to the radio data for M82. We tested several models and found that the free-free covering factor was always near unity. A simpler model consisting of a cylindrical wall of warm, ionized gas surrounding a lower density hot medium provides a reasonable representation of the free-free radio absorption in M82.

The multiphase nature of the ISM is incorporated by considering the different properties of the warm and hot ionized gas. The warm, ionized gas is assumed to be magnetically threaded and produces both synchrotron and free-free emission ($j_{WIM} = j_{\nu}^{synch} + j_{\nu}^{ff,ion}$) as well as free-free absorption ($\kappa_{WIM} = \kappa_{\nu}^{ff,ion}$). Synchrotron self-absorption is found to be negligible in M82, so we need only consider free-free absorption.  The hot, low density gas also produces synchrotron, and due to its lower density, negligible amounts of free-free emission and absorption ($j_{hot} = j_{\nu}^{synch} + j_{\nu}^{ff,hot} \approx j_{\nu}^{synch}$).

Adopting our assumed cylindrical geometry with a shell of warm, ionized gas, we take the shell width to be $l/2$ where $l$ is the radius of the cylinder.  Because the galaxy is nearly edge-on, we see emission from both the front and the back of the cylinder.  Emission from the front of the cylinder is
\begin{equation}\label{ionized1}
I_{\nu, front} = \frac{j_{WIM}}{\kappa_{WIM}} \left( 1 - e^{- \kappa_{WIM} l/2} \right).
\end{equation}
Emission from the rear of the cylinder can be absorbed within both the front and back warm ionized gas shells. Thus,
\begin{equation}\label{ionized}
I_{\nu, back} = \frac{j_{WIM}}{\kappa_{WIM}} \left( 1 - e^{- \kappa_{WIM} l/2} \right) e^{- \kappa_{WIM} l/2}.
\end{equation}
Then, the radiative intensity emergent from the hot, diffuse gas is given by
\begin{equation}\label{diffuse}
I_{\nu, hot} = 2 j_{hot} r_{i} e^{- \kappa_{WIM} l/2},
\end{equation}
where $r_{i}$ is the radius of the central volume of hot, diffuse gas ($r_{starburst} = r_{i} + l/2$).  Then, the total emergent intensity from the starburst region is 
\begin{equation}
I_{\nu} = \frac{j_{WIM}}{\kappa_{WIM}} \left( 1 - e^{- \kappa_{WIM} l} \right) + 2 j_{hot} r_{i} e^{- \kappa_{WIM} l/2}.
\end{equation}

This model ignores the presence of a radio halo.  Observations of the radio halo shows it has a steeper spectral index than that of the main body and thus can become dominant at low frequencies \citep{Seaquist91,Adebahr}.  We therefore constructed a model that assumes all of the 74~MHz flux from M82 arises in an unabsorbed synchrotron halo.  This assumption is consistent with the parameters of our cylindrical model, where any 74~MHz emission from the main starburst zone is heavily absorbed. We assume a standard -0.7 spectral index for the halo.  The radio halo's spectral energy distribution then is simply
\begin{equation}\label{halo}
I_{\nu, halo} = I_{0, halo} \nu^{-0.7},
\end{equation}
where $I_{0, halo}$ is determined by our lowest frequency radio data point.  The final expression for intensity is
\begin{equation}
I_{\nu} = \frac{j_{WIM}}{\kappa_{WIM}} \left( 1 - e^{- \kappa_{WIM} l} \right) + 2 j_{hot} r_{i} e^{- \kappa_{WIM} l/2} + I_{\nu, halo}.
\end{equation}

Figure 6 shows the radio spectrum predicted by the simple cylindrical model, which shows the contributions from the model components.  Emission from the hot, diffuse gas, eqn. (\ref{diffuse}), is represented by the dotted blue line; emission from the warm, ionized gas, eqn. (\ref{ionized1}) \& (\ref{ionized}), is represented by the dot-dashed green line; and emission from the halo, eqn. (\ref{halo}), is represented by the dashed red line.  Note that the free-free high frequency emission is under estimated since we cannot use a single value for the mean electron density (see \S 4.4). Our best-fit model predicts a flux from the starburst region at 333~MHz in agreement with direct measurements from \cite{Adebahr}.

\subsubsection{Radio Observations}

\begin{figure*}[t!]
\epsscale{1.15}
\plottwo{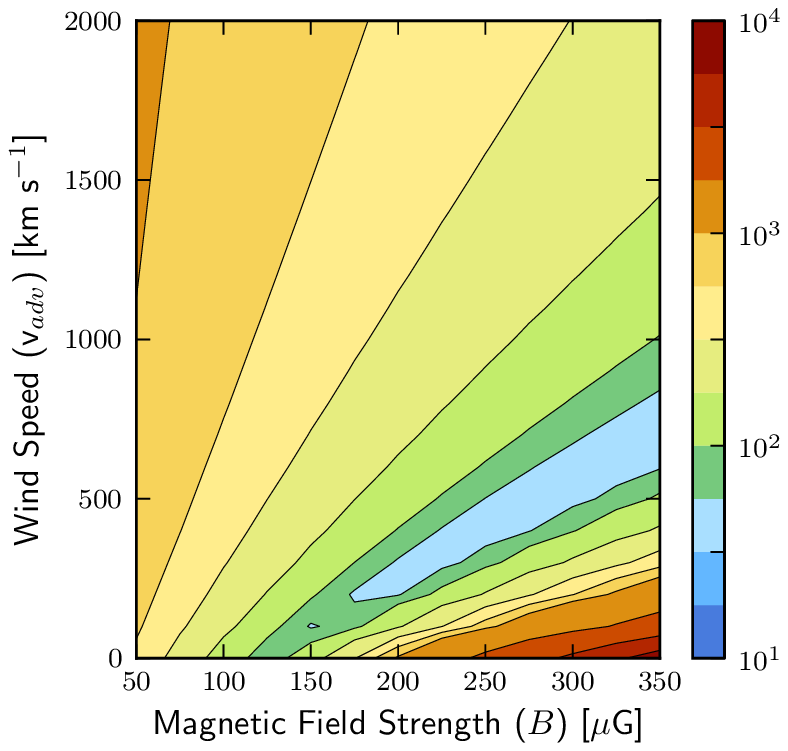}{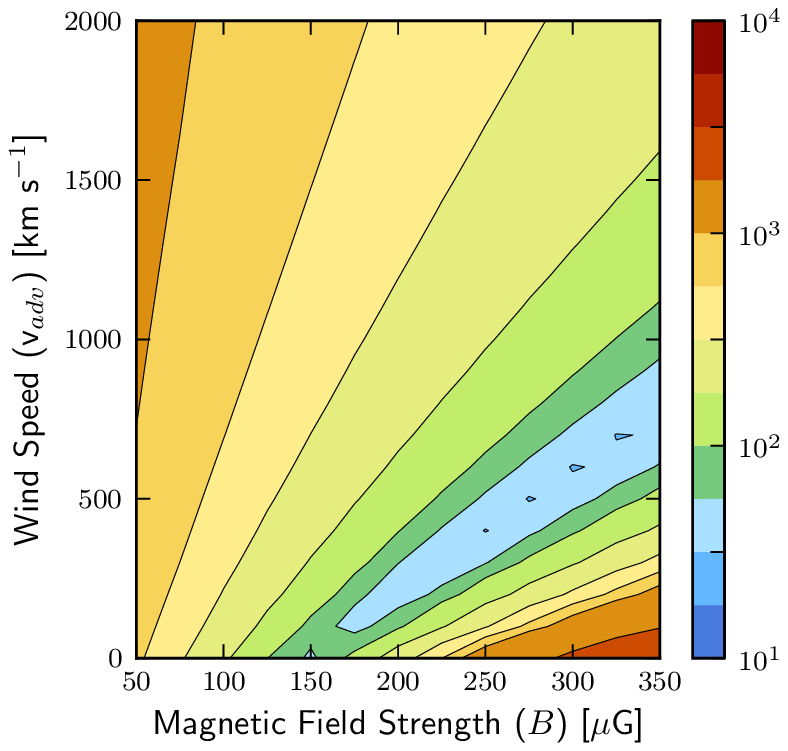}
%
%
\plottwo{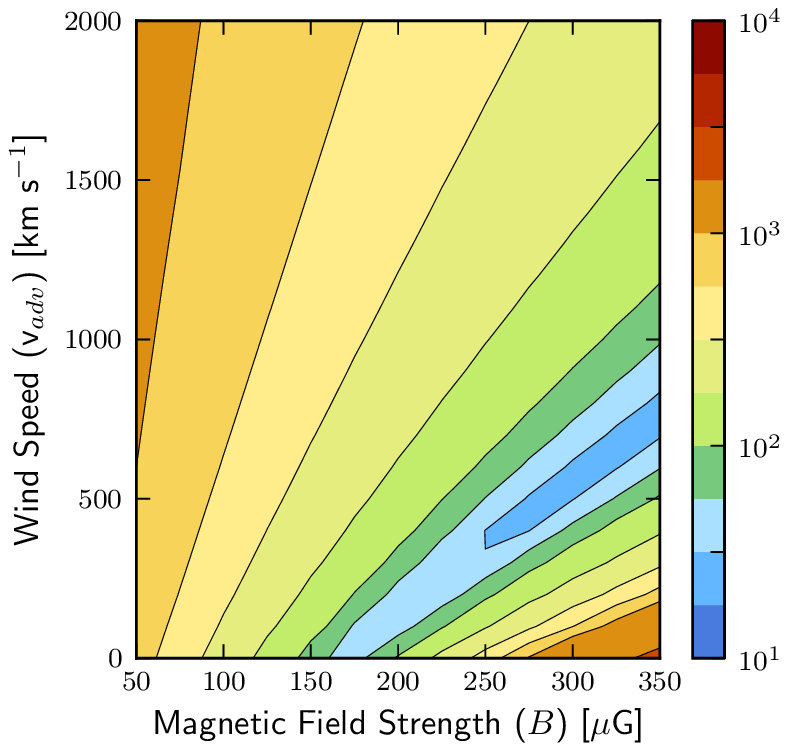}{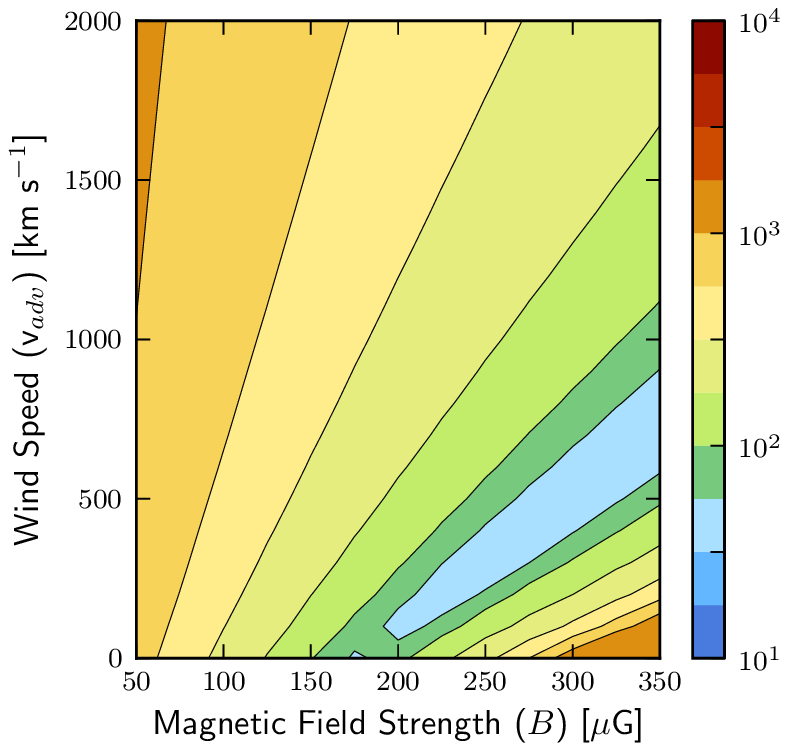}
\caption{Contour plots showing changes in $\chi^{2}$ values for changes in magnetic field strength ($B$) and advection (wind) speed ($v_{adv}$).  \textit{Top left:} Parameters set at $p$ = 2.1, $M_{mol} = 2 \times 10^{8}$~M$_{\odot}$, $n_{ion}$ = 300~cm$^{-3}$.  \textit{Top right:} Parameters set at $p$ = 2.1, $M_{mol} = 3 \times 10^{8}$~M$_{\odot}$, $n_{ion}$ = 150~cm$^{-3}$.  \textit{Bottom left:} Parameters set at $p$ = 2.1, $M_{mol} = 4 \times 10^{8}$~M$_{\odot}$, $n_{ion}$ = 100~cm$^{-3}$.  \textit{Bottom right:} Parameters set at $p$ = 2.2, $M_{mol} = 4 \times 10^{8}$~M$_{\odot}$, $n_{ion}$ = 150~cm$^{-3}$.}
\end{figure*}

Because of the differing methods of observation for the \cite{Klein88} and \cite{Williams10}, data sets it is difficult to say which is more accurate.  At most, we can say that the single-dish observations serve as an upper limit while the interferometer observations provide a lower limit for the radio fluxes for the starburst region of M82.  For comparison with our model, we calculated the median between the two data sets and scaled down the data from \cite{Klein88} as it encompasses a larger range of frequencies than the \citeauthor{Williams10} data.  We also increased the error bars to six percent for the \citeauthor{Klein88} data to include both data sets.

We do not include the single-dish 87.2~GHz measurement of \cite{Jura78}, included in the \citeauthor{Klein88} data set, as it is unusually low compared to even the interferometer data \citep{Klein88, Williams10}.  Instead, we use the 92~GHz interferometer measurement from \cite{Carlstrom91}.  Additionally, we include newer low frequency measurements at 74~MHz \citep{Cohen07} and 333~MHz \citep{Basu12}.  As these are both interferometer observations, we scale them up to use with our median data set.  Distinguishing between core and halo emission at these low frequencies is quite difficult \citep{Adebahr} and so we incorporate a halo emission component into our model (see above).

\subsection{Results}

\subsubsection{$\chi^{2}$ Tests}

\begin{center}
\begin{deluxetable}{llc}
%
\tablecaption{Fitted Model Parameters}
\tablewidth{0pt}
\tablehead{
\colhead{Physical Parameters} & \colhead{Tested Range} & \colhead{Reference}
}
\startdata
Magnetic Field Strength ($B$) & 50-350 $\mu$G & 1,2\\
Advection (Wind) Speed ($v_{adv}$) & 0-2000 km~s$^{-1}$ & 3,4\\
Ionized Gas Density ($n_{ion}$) & 50-500 cm$^{-3}$ & 5\\
\enddata
%
%
\tablerefs{
[1] \cite{delPozo09a}; [2] \cite{Thompson06}; [3] \cite{Shopbell98}; [4] \cite{Strickland09}; [5] \cite{Forster01};
}
\end{deluxetable}
\end{center}

One of the primary goals of developing a one-zone model is to be able to find a set of parameters which produce an acceptable fit to both the radio and $\gamma$-ray spectra.  To find the best-fit combined solutions, we use $\chi^{2}$ tests on both spectra.  We begin by using observations and previous studies to determine the range of values to test for each parameter.  Previous studies have calculated the equipartition magnetic field strength to be $\sim$150~$\mu$G \citep{delPozo09a}.  Thus, we test values from 50 to 350 $\mu$G.  For the wind speed, model fits to x-ray observations of the wind give wind speeds on the order of 1500 km~s$^{-1}$ \citep{Strickland09}, while optical observations give wind speeds on the order of 500 km~s$^{-1}$ \citep{Shopbell98}.  The two measurements do not inherently contradict each other as they mesaure different materials in the wind.  Thus, we test values from 0 to 2000 km~s$^{-1}$.  We have fixed the total mass of the molecular and ionized gases, and in doing so, we have also fixed the proton density.  Observations constrain the ionized gas densities to be 10 to 600~cm$^{-3}$ with kinetic temperatures of $\sim$8000 K \citep{Forster01}.  We test values from 50 to 500~cm$^{-3}$ (see Table 2).  Additionally, we test two different cosmic ray spectral indices ($p$ = 2.1, 2.2) and three different molecular gas masses ($M_{mol}$ = $2 \times 10^{8}$, $3 \times 10^{8}$, $4 \times 10^{8}$ M$_{\odot}$).  We took advantage of the computational simplicity of our model to run 49,000+ cases.

\begin{figure*}[t!]
\epsscale{0.9}
\plotone{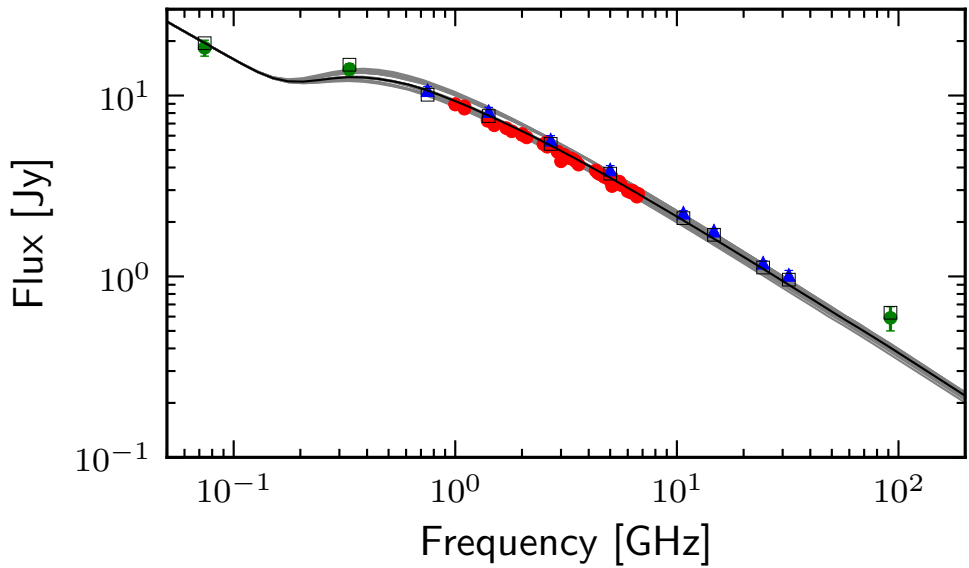}
\caption{All fits within 3$\sigma$ of minimum $\chi^{2}$ for $n_{ion}$ = 100~cm$^{-3}$ with $p$ = 2.1, $M_{mol}$ = $4 \times 10^{8}$~M$_{\odot}$.  The black line represents the best-fit ($\chi^{2}$ = 22.6) and the grey lines represent all other 3$\sigma$ fits.  Radio data include \cite{Klein88} (blue triangles), \cite{Williams10} (red circles), \cite{Cohen07} (green circle), \cite{Basu12} (green circle), \cite{Carlstrom91} (green circle).  Open squares represent the median data set used for the $\chi^{2}$ tests.}
\end{figure*}

In determining the best fit to the radio spectrum, we performed $\chi^{2}$ tests using the available radio observations, discussed above.  Fits to the eleven radio data points were used to constrain the ionized gas density ($n_{ion}$), the advection speed ($v_{adv}$), and the magnetic field strength ($B$) (see Figure 7).  For the $\gamma$-rays, fits to the eight data points (excluding upper limits) were used to place additional constraints on the advection speed and the magnetic field strength.

In looking for the joint best-fit model, we found that for the radio spectrum there is a degeneracy between magnetic field strength and wind speed such that as wind speed increases, an increase in magnetic field strength allows a similiarly good fit (see Figure 7).  We explain the reasons for this degeneracy below.  For the $\gamma$-ray spectrum we also find a valley of solutions (several minima as opposed to a single absolute minimum).  If we restrict the possible $\gamma$-ray models to those deemed to have acceptable radio spectra, we find that the $\chi^{2}$ tests for the $\gamma$-ray spectrum limits the tested wind speeds to only a few.

The best-fit solution has a $\chi^{2}$ value of $\chi^{2} = 22.6$ for the radio spectrum and $\chi^{2} = 9.6$ for the $\gamma$-ray spectrum.  It occurs for an ionized gas density of 100 cm$^{-3}$, a magnetic field strength of $B= 275$ $\mu$G, and a wind speed of $v_{adv} = 500$ km~s$^{-1}$ (Table 3).  The models within 3$\sigma$ have magnetic field strengths of $B$ = 225-350 $\mu$G, advection speeds of $v_{adv}$ = 300-700 km~s$^{-1}$, and ionized gas densities of $n_{ion}$ = 50-250 cm$^{-3}$.  Figure 8 shows all radio spectra for models within 3$\sigma$ of the minimum for a single spectral index and molecular gas mass.  Of the 16,000+ models tested, we found 666 radio models within 3$\sigma$ while only 89 had both radio and $\gamma$-ray models within 1$\sigma$.

The best-fit $\gamma$-ray model can be seen in Figure 5 and the corresponsing radio model can be seen in Figure 6.  The radio spectrum agrees to within the errors for ten of eleven data points, while the $\gamma$-ray spectrum agrees with six of eight data points.  It can be clearly seen in Figure 6 that the radio model does not agree with the 92 GHz data point.  The flux at 92 GHz has been previously shown to be mostly due to free-free emission \citep{Williams10}.  Because we have limited the ionized gas to a single density, we achieve the turnover at low frequencies due to free-free absorption but not the required flux at high frequencies due to free-free emission.  To achieve both the observed free-free absorption and emission would require another level of detail to the model.  Thus, in maintaining the simplicity of the model, we are not able to produce the required free-free flux at high frequencies.

\begin{center}
\begin{deluxetable}{llc}
%
\tablecaption{Best-Fit Model Parameters}
\tablewidth{0pt}
\tablehead{
\colhead{Physical Parameters} & \colhead{Best-Fit Value}
}
\startdata
Magnetic Field Strength ($B$) & 275 $\mu$G\\
Advection (Wind) Speed ($v_{adv}$) & 500 km~s$^{-1}$\\
Ionized Gas Density ($n_{ion}$) & 100 cm$^{-3}$\\
Spectral Index ($p$) & 2.1\\
Molecular Gas Mas ($M_{mol}$) & $4 \times 10^{8}$ M$_{\odot}$\\
\enddata
\tablecomments{Results for $\chi^{2}_{radio}$ = 22.6, $\chi^{2}_{\gamma}$ = 9.6}
\end{deluxetable}
\end{center}
%

\subsubsection{Wind Speed and Magnetic Field Strength}

In the course of our $\chi^{2}$ tests, we found a degeneracy between the magnetic field strength and the wind speed.  As one increases the magnetic field strength, you must also increase the wind speed to achieve a similarly good fit.  Why might such a degeneracy occur?  

\begin{figure*}[t!]
\epsscale{1.15}
\plottwo{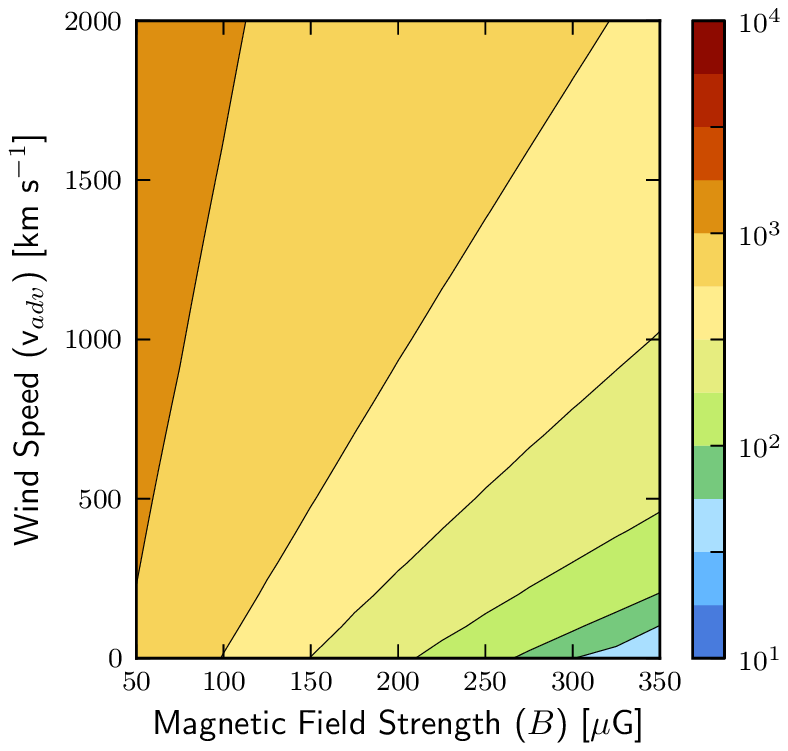}{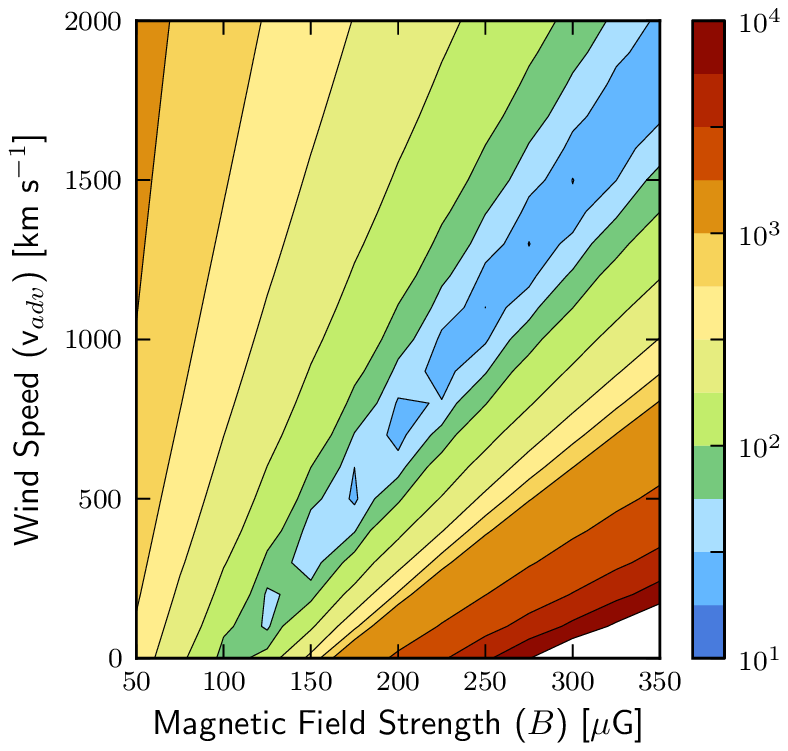}
\caption{Contour plots showing changes in $\chi^{2}$ values for changes in magnetic field strength ($B$) and advection (wind) speed ($v_{adv}$) for acceleration efficiencies of 4\% and 20\%.  \textit{Left:} Parameters set at $p$ = 2.1, $M_{mol} = 4 \times 10^{8}$~M$_{\odot}$, $n_{ion}$ = 150~cm$^{-3}$, $\eta$ = 0.04.  \textit{Right:} Parameters set at $p$ = 2.1, $M_{mol} = 4 \times 10^{8}$~M$_{\odot}$, $n_{ion}$ = 50~cm$^{-3}$, $\eta$ = 0.2.}
\end{figure*}

The magnetic field strength and the wind speed both affect the amount of synchrotron emission.  The magnetic field essentially scales the radio spectrum up and down as the square of the magnetic field strength; however, because the high frequency portion of the spectrum is dominated by thermal emission from the ionized gas, changing the magnetic field does not change the high frequency end.  If we increase the wind speed, then we decrease the advection time and thus decrease the number of cosmic-ray electrons and positrons in the starburst region.  By advecting away some of the electrons/positrons, we decrease the amount of synchrotron emission.  Because we have fixed the molecular and ionized gas masses and therefore the average density, it is necessary to have advection to create a better fit as otherwise we overpredict the radio spectrum.  If we increase the wind speed too much, then the advection time is shorter than the loss lifetime, and we no longer have enough cosmic-ray protons to produce the correct amount of $\gamma$-ray flux.

\subsubsection{Gas Mass and Spectral Index}

In addition to varying the values for ionized gas density, magnetic field strength, and wind speed, we also tested different values for spectral index and molecular gas mass.  We test a much smaller range of values for the cosmic ray injection spectral index ($p$ = 2.1, 2.2) and molecular gas mass ($M_{mol}$ = $2 \times 10^{8}$, $3 \times 10^{8}$, $4 \times 10^{8}$ M$_{\odot}$) than for the other parameters.  While the observations relating to ionized gas density, magnetic field strength, and wind speed leave much uncertainty as to their true values, observations for spectral index and molecular gas mass are more constraining.  Thus, we only need test a few values for each variable.

We find that for both the radio and $\gamma$-ray spectra, the spectral index of $p$ = 2.1 is a better fit than $p$ = 2.2.  While selecting $p$ = 2.1 over $p$ = 2.2 leads to less than a factor of 2 reduction in the minimum $\chi^{2}$ value, the $\gamma$-ray models with  $p$ = 2.2 consistently miss the TeV VERITAS data points (see Figure 5).  For molecular gas mass, we find that the minimum $\chi^{2}$ value decreases as the gas mass increases with there being a slightly larger gap between in $\chi^{2}$ values between $2 \times 10^{8}$ and $3 \times 10^{8}$ than between $3 \times 10^{8}$ and $4 \times 10^{8}$.


\section{Discussion}

Over the course of this paper, we have investigated a model for cosmic ray interactions in the starburst galaxy M82.  We incorporate synchrotron emission and free-free absorption and emission into the radio spectrum and include neutral pion decay and bremsstrahlung as mechanisms for $\gamma$-ray production.  For the most part, the fits agree well with the data, with the exception of 92 GHz radio data point.  The radio fits also indicate that there is a degeneracy between magnetic field strength and wind speed.  We find that for both the radio and $\gamma$-ray spectra a spectral index of $p$ = 2.1 and a molecular gas mass of $M_{mol} = 4 \times 10^{8}$ M$_{\odot}$ produce the best fit (see \S 4.4), but a variety of different models fit both the radio and $\gamma$-ray spectra well.  However, while the $\chi^{2}$ tests on the radio spectrum allow for fits over a large range of wind speeds, the $\chi^{2}$ tests on the $\gamma$-ray spectrum more severely limit the range of allowable wind speeds.

\subsection{Is M82 a Calorimeter?}

One of the main questions we seek to answer in this paper is whether or not M82 is a calorimeter.  We define a calorimeter such that cosmic-ray proton or electron energy losses dominate over advective or diffusive losses.  In the case of the cosmic-ray electrons, it is clear that the galaxy is an electron calorimeter as energy losses dominate over advective losses for all wind speeds tested (see Figure 1).

However, it is more difficult to decide exactly when a galaxy is no longer a proton calorimeter.  In the case of the cosmic-ray protons, the advection timescale is about equal to the peak of the energy loss lifetime (see Figure 1) meaning that the galaxy is a 50\% calorimeter.  How calorimetric a galaxy is depends on not only on the wind speed, which determines the advection timescale, and the average density, which determines the energy loss timescale, but also on the cosmic ray acceleration efficiency.  Throughout the paper, we have assumed cosmic ray acceleration efficiency ($\eta$) of 10\%.  To investigate the impact of acceleration efficiency, we tested two alternative values: a decreased efficiency of $\eta = 4$\% and an increased efficiency of $\eta = 20$\% (that could also be achieved with an efficiency of 10\% and an supernova rate of 0.14 yr$^{-1}$).

For the 4\% case, the resulting radio models consistently required higher magnetic field strengths due to a lack of electrons because of the decrease in the primary electron population and because of a decrease in the secondary electron/positron population because of the decrease in primary protons.  The best models required wind speeds on the order of 0-100~km~s$^{-1}$ (see Figure 9).  Wind speeds of this magnitude give advection timescales that are longer than the proton energy loss timescales, resulting in an 80\%-100\% calorimeter.  The best fit had a $\chi^{2}$ value nearly a factor of 2 larger than the best fit for an efficiency of $\eta = 10$\% for both the $\gamma$-ray and radio spectra.  For all three acceleration efficiencies, we found 317 of the 49,000+ models to be within 1$\sigma$ for both the radio and $\gamma$-ray spectra.  Of those 317 models, only 45 had an acceleration efficiency of 4\%.  Based on these results and the extremely low wind speeds, we can effectively rule out the 4\% case.

Conversely, for the 20\% efficiency case, we found that significantly higher wind speeds were required to achieve good fits because of an increase in the overall electron/positron population.  The best models required wind speeds on the order of $\sim$1500~km~s$^{-1}$ (see Figure 9).  While models for an efficiency of 10\% agree with measurements of wind speeds in the optical \citep{Shopbell98}, wind speeds on the order of 1500~km~s$^{-1}$ agree with x-ray measurements by \cite{Strickland09}.

Comparing the resulting advection timescales to the proton energy loss timescales gives a 35\% calorimeter.  We find that for this higher acceleration efficiency, the minimum $\chi^{2}$ value for the radio spectrum is smaller than in the 10\% case, while the $\chi^{2}$ value for the $\gamma$-ray spectrum is just slightly higher than before ($\chi^{2} = 10.8$ vs $\chi^{2} = 9.6$).  Thus, while we rule out the 4\% acceleration efficiency, we consider the 20\% case to be plausible.

Acceleration efficiency directly affects our initial cosmic ray population.  However, it is not the only factor.  For example, we use the lower bound of 0.07 SN~yr$^{-1}$ for our supernova rate.  If we scaled the supernova rate up and the acceleration efficiency down in equal measure, our initial population of cosmic rays would be unchanged.  While we have selected one particular set of parameters for our initial cosmic ray population, it is simple to obtain the same population from a slightly different set of initial parameters and as such, we would again produce a degenerate relationship between magnetic field strength and wind speed.  Thus, we would produce the same fits as presented above.

\subsection{FIR-Radio Correlation}

It remains a curious question as to why starburst galaxies lie on the same FIR-radio correlation as the normal, quiescent galaxies.  The general explanation for the FIR-radio correlation in disk galaxies is the electron calorimeter theory by \cite{Volk89}.  This paper postulates that both FIR and radio synchrotron emission are proportional to the supernova rate as FIR emission is essentially starlight reradiated by dust grains and radio synchrotron emission comes from energetic electrons accelerated by supernovae \citep{Volk89}.  As such, it is required that cosmic-ray electrons radiate away their energy wihtin the galaxy to achieve the observed radio synchrotron emission.  However, in dense environments such as starburst galaxies, there is a second source of electrons/positrons from proton-proton collisions.

The overall importance of secondary electrons/positrons is dependent upon the ratio of primary protons to primary electrons ($N_{p}/N_{e}$).  The significance of secondary electron/positrons decreases as the ratio of primary protons to electrons decreases.  In our model, we leave this ratio as an independent free parameter fixed at 50 based upon supernovae studies in the Milky Way.  \cite{Schlick02} makes a case for this parameter being model dependent such that the ratio depends on the masses of protons and electrons and the selected spectral index, $N_{p}/N_{e} = (m_{p}/m_{e})^{(p-1)/2}$.  Our prefered spectral index is $p = 2.1$, which gives a ratio of $N_{p}/N_{e} \sim 60$.  As our free parameter selection is smaller than this value, the importance of secondary electrons/positrons would only increase by making the ratio model dependent.

Thus, while we find that M82 is an electron calorimeter, the overall picture is complex.  It is true that all cosmic ray electrons are mainly confined to the starburst core.  However, because the galaxy is not also a proton calorimeter, there is an inherent loss of any secondary electrons and positrons that would have resulted from proton-proton collisions in the core.  Thus, any electrons produced in the starburst region (primary or secondary) are confined to the core but the total possible population of electrons/positrons is not equivalent to the confined population.  This is one of the mechanisms responsible for keeping the galaxy on the FIR-radio correlation.  Without some loss of protons to a advection in a galactic wind, there would be a larger population of electrons/positrons resulting in a higher radio flux.

\subsection{Comparison with Other Models}

Our approach to modeling cosmic rays in starburst galaxies builds on that taken in previous models \citep{delPozo09a, Paglione12}.  However, we place an emphasis in our models on varying and constraining a variety of possible wind (advection) speeds rather than assuming a single, static wind speed \citep{delPozo09a, Paglione12}.  Additionally, whereas some previous models have had more free parameters \citep{Lacki10}, our approach reduces the varied parameters to a select few.  All other parameters are chosen to be in agreement with observations.  Thus, we compare large numbers of models and strive to produce fits to the radio and $\gamma$-ray spectra that are more intuitive and observationally informed.

\cite{Paglione12} favor higher mean gas densities, ranging from 100 to 1000~cm$^{-3}$, and strong magnetic fields, ranging from 200 to 1000~$\mu$G.  Their best solution resulted in a density of $n$ = 600~cm$^{-3}$ and a magnetic field strength of $B$ = 450~$\mu$G.  In contrast, we favor smaller magnetic field strengths (225 to 350~$\mu$G).  However, though we test smaller densities (280 to 415~cm$^{-3}$), our best fits have a comparable density of 550~cm$^{-3}$ (see Table 1).  \cite{delPozo09a} favor models with larger supernova rates (0.1-0.3 SN~yr$^{-1}$) and use a significantly smaller average ISM density (180~cm$^{-3}$).  They produce models with magnetic fields strengths from 120 to 290~$\mu$G.  This range is comparable to our own results.

Like our model, both \cite{delPozo09a} and \cite{Paglione12} incorporate free-free absorption into their radio models with spectra that turn over at $\sim$700~MHz and $\sim$200~MHz respectively.  Neither incorporates a halo synchrotron emission component to allow for comparison with low frequency radio data.  Additionally, \cite{Paglione12} assume a constant convection loss timescale of 1 Myr which is sufficiently longer than the energy loss timescale for cosmic rays (at their preferred average density of 600 cm$^{-3}$) that they find the galaxy to be calorimetric for both protons and electrons.

In addition to creating a model for cosmic rays in starburst galaxies, we address the question of calorimetry in starburst galaxies.  Calorimetry is also address by the recent paper \cite{Ackermann12}.  A semi-analytical formula which depends on the star-formation rate and supernova acceleration efficiency of a galaxy is used to determine the calorimetric efficiency estimate, while we compare a fully calculated energy loss lifetime against the advection timescale for a variety of wind speeds (see \S 3.1 and \S 5.1).  \cite{Ackermann12} find that galaxies with SFR $\sim10$ M$_{\odot}$~yr$^{-1}$ have calorimetric efficiencies of 30\% to 50\%, a similar result to our own findings.


\section{Conclusions}

While studying the complex situation presented by cosmic ray interactions in the starburst galaxy M82, we found it useful to develop a physically realistic but computationally simple single zone model.  In this paper, we present a cosmic ray model for M82 that has few free parameters and is observationally informed.  Our model produces statistically reasonable fits to both the radio and $\gamma$-ray spectra.  We find that the inclusion of secondary electrons and positrons as well as a galactic wind are key to producing better fits.

In the course of modeling the radio spectrum, we find that geometry is vital due to a rather high ionized gas optical depth at low frequencies.  We achieve the turnover of the radio spectrum in the starburst core due to free-free absorption, but do not reproduce the observed high frequency free-free emission.  Without adding another level of detail, it is impossible to achieve both the spectrum turnover at low frequencies due to free-free absorption and the necessary flux from free-free emission at high frequencies.

While M82 is undoubtedly an electron calorimeter, it is not straightforward to make such a distinction for the cosmic-ray protons.  With our original assumption of an acceleration efficiency of 10\%, M82 is a 50\% proton calorimeter, meaning energy losses within the starburst zone equal advection losses in the galactic wind.  We found that it was not possible to achieve acceptable fits for an 80\% calorimeter when decreasing the acceleration efficiency to 4\%.  However, of 49,000+ models, fits can be achieved for an increase of the cosmic ray acceleration efficiency to 20\% (or with an efficiency of 10\% and a supernova rate of 0.14 yr$^{-1}$) such that M82 is a 35\% proton calorimeter.  Thus, M82 is at best a 50\% calorimeter and from these results, it is clear that a wind is vital to the modeling of cosmic ray interactions in the starburst region.

One of the implications of the FIR-radio correlation is that the magnetic and cosmic-ray energy density scale with the SFR.  From our models, we find that this is partially fulfilled in M82.  In particular, our $\chi^{2}$ analysis puts the magnetic field near (slightly above) equipartition, independent of any minimum energy assumption.  This is also fulfilled in that we find M82 to be an electron calorimeter.  Thus, our results go part way toward explaining the FIR-radio correlation.  However, as we do not find the galaxy to be a proton calorimeter, the cosmic-ray energy density does not completely scale with the SFR.  As a consequence of not being a perfect proton calorimeter, both the $\gamma$-ray and radio spectra are dimmer than they would be otherwise, as the loss of protons leads to a loss of secondary pions and thus a loss of both $\gamma$-rays and secondary positrons and electrons.  Additionally, the neutrino flux will be a factor of a few less than predicted by calorimeter models.


\acknowledgements

This work was supported in part by NSF AST-0708967, NSF AST-0903900, and NSF PHY-0821899 (to the Center for Magnetic Self-Organization in Laboratory and Astrophysical Plasmas) and NSF PHY-0969061 (to the IceCube Collaboration).  We thank Ralf-J\"{u}rgen Dettmar, Julia Becker Tjus, Reinhard Schlickeiser, and Bj\"{o}rn Adebahr for discussions and hospitality during a visit to the Ruhr-Universit\"{a}t Bochum.  We also thank Francis Halzen for his help and support.


\appendix


\textit{Energy loss rates:}  For protons, losses include ionization, Coulomb effects, and pion production.  Ionization losses, for a neutral medium, are proportional to the density of the medium and weakly dependent on energy such that
\begin{equation}
- \left( \frac{dE}{dt} \right)_{Ion,p} = 1.82 \times 10^{-7} ~ n \left(1 + 0.0185 ~ ln(\beta) ~ \Theta[\beta - 0.01]\right) \frac{2 \beta^{2}}{10^{-6} + 2 \beta^{3}} ~~ \text{eV~s$^{-1}$},
\end{equation}
where $\beta = v/c$ and $\Theta$ denotes the heavyside theta function \citep{Schlick02}.  In an ionized medium, Coulomb losses depend on the electron density ($n_{e}$) and temperature ($T_{e}$) such that \citep{Schlick02}
\begin{align}
- \left( \frac{dE}{dt} \right)_{Coul,p} = ~ &3.08 \times 10^{-7} ~ n_{e} \frac{\beta^{2}}{\beta^{3} + 2.34 \times 10^{-5} \left( T_{e}/2 \times 10^{6} K \right)^{3/2}} \nonumber \\ &\times \Theta \left[\beta - 7.4 \times 10^{-4} \left(\frac{T_{e}}{2 \times 10^{6} K} \right)^{1/2} \right]  ~~ \text{eV~s$^{-1}$}.
\end{align}
The dominant loss for protons above $\sim$1 GeV is pion production.  The threshold proton energy for pion production is 1.22 GeV \citep{Schlick02} and the energy loss rate is given by
\begin{equation}
- \left( \frac{dE}{dt} \right)_{Pion,p} = 1.31 \times 10^{-7} ~ n ~ \gamma^{1.28} ~ \Theta[\gamma - 1.3] ~~ \text{eV~s$^{-1}$}.
\end{equation}

For electrons, losses include ionization, bremsstrahlung, inverse Compton, and synchrotron.  Ionization losses are weakly dependent on energy and are proportional to density such that
\begin{equation}
- \left( \frac{dE}{dt} \right)_{Ion,e} = 36.18 ~ c \sigma_{T} n (1 + 0.146 ~ ln(\gamma) + 1.54 ~ n_{e}) ~~ \text{eV~s$^{-1}$}.
\end{equation}
where $\sigma_{T}$ is the Thomson cross section, $sigma_{T} = 6.65 \times 10^{-25}$ cm$^{2}$.  Bremsstrahlung is the dominant energy loss below $\sim$few GeV.  Energy losses due to bremsstrahlung are proportional to energy and dependent on the medium density and the energy loss rate is given by
\begin{equation}
- \left( \frac{dE}{dt} \right)_{Brem,e} = n E \frac{Z^{2} e^{6}}{12 \pi^{3} \epsilon_{0}^{3} m_{e}^{2}c^{4} h} ln\left( \frac{192}{Z^{1/3}} \right) ~~ \text{eV~s$^{-1}$},
\end{equation}
where $Ze$ is the charge of the nucleus \citep{Longair11}.  Inverse Compton scattering goes as the square of the electron energy and depends on the radiation field energy density, $U_{rad}$, such that \citep{Schlick02}.
\begin{equation}
- \left( \frac{dE}{dt} \right)_{IC,e} = \frac{4}{3} ~ c ~ \sigma_{T} \left( \frac{E}{m_{e}c^{2}} \right)^{2} U_{rad} ~~ \text{eV~s$^{-1}$}.
\end{equation}
The synchrotron energy loss rate also depends on the square of energy and the magnetic field energy density such that \citep{Schlick02}
\begin{equation}
- \left( \frac{dE}{dt} \right)_{Synch,e} = \frac{4}{3} ~ c ~ \sigma_{T} \left( \frac{E}{m_{e}c^{2}} \right)^{2} \frac{B^{2}}{2 \mu_{0}} ~~ \text{eV~s$^{-1}$}.
\end{equation}
Synchrotron and inverse Compton losses are dominant above a few GeV.


%
\end{document}